\documentclass[10pt,english,twocolumn]{revtex4-2}
\usepackage[T1]{fontenc}
\usepackage[utf8]{inputenc} 
\usepackage{geometry}
\usepackage{units}
\usepackage{textcomp}
\usepackage{amsmath}
\usepackage[english]{babel}
\usepackage{graphicx}
\usepackage{xcolor}
\usepackage[version=4]{mhchem} 

\begin{document}



















\author{E. Billaud$^1$, L. Balembois$^1$, J. Travesedo$^1$, M. Le Dantec$^1$, M. Ran\v{c}i\'{c}$^1$, E. Albertinale$^1$, R. Truong$^1$, S. Bertaina$^2$, T. Chaneli\`ere$^3$, P. Goldner$^4$, D. Est\`eve$^1$, D. Vion$^1$, E. Flurin$^1$, P. Bertet$^{1}$}

\email{patrice.bertet@cea.fr}

\affiliation{$^1$Universit\'e Paris-Saclay, CEA, CNRS, SPEC, 91191 Gif-sur-Yvette Cedex, France\\
$^2$CNRS,  Aix-Marseille  Universit\'e,  IM2NP  (UMR  7334),  Institut  Mat\'eriaux Micro\'electronique  et  Nanosciences de Provence,  Marseille,  France \\
$^3$Univ. Grenoble Alpes, CNRS, Grenoble INP, Institut N\'eel, 38000 Grenoble, France \\
$^4$Chimie ParisTech, PSL University, CNRS, Institut de Recherche de Chimie Paris, 75005 Paris, France}

\title{Electron Paramagnetic Resonance spectroscopy of a scheelite crystal using microwave photon counting}

\begin{abstract}
Counting the microwave photons emitted by an ensemble of electron spins when they relax radiatively has recently been introduced as a sensitive new method for electron paramagnetic resonance spectroscopy at millikelvin temperatures. Here, we apply this spin fluorescence method to a scheelite crystal of CaWO4, finding some known  ($\mathrm{Er}^{3+}$, $\mathrm{Yb}^{3+}$, $\mathrm{Nd}^{3+}$ and $\mathrm{Fe}^{3+}$) and other unknown paramagnetic impurities. Investigating the zero nuclear spin isotope ($I=0$) transition of $\mathrm{Er}^{3+}:\mathrm{CaWO}_4$ as a model system, we provide a quantitative analysis of the time-dependent photon counting rate following an excitation pulse, as a function of its power. The achieved signal-to-noise ratio is found to be an order of magnitude higher than the one obtained by inductively-detected Hahn echo under identical conditions. Finally, we use spin fluorescence spectroscopy at low excitation power to probe the properties of rare-earth-ions close to a metallic wire deposited on the surface; our data reveal line distortion caused by the mechanical strain imparted by the thermal contractions of the metal relative to the underlying crystal. Coherent oscillations are also observed for the most highly strained ions.
\end{abstract}

\maketitle

\section{Introduction}


Electron Paramagnetic Resonance (EPR) spectroscopy is an important technique for characterizing the density and properties of paramagnetic centers in a sample. In the usual Inductive-Detection (ID) scheme, the sample is inserted into a microwave resonator with angular frequency $\omega_0$ and probed by sequences of microwave pulses, a prominent one being the Hahn echo sequence $\pi/2 - \tau - \pi - \tau - \mathrm{echo}$. When the transition frequency of the paramagnetic impurities is tuned to $\omega_0$ by the application of a static magnetic field $B_0$, a transient macroscopic transverse magnetization is built up at a time $\tau$ after the $\pi$ pulse, whose oscillation at the electronic Larmor frequency leads to the emission of a coherent microwave pulse called a spin echo. Measuring the echo amplitude as a function of $B_0$ gives access to spectroscopic information such as spin density and line shape and width. Together with measurements of transient spin-coherence and various other hyperfine spectroscopic techniques~\cite{schweiger_principles_2001}, EPR spectroscopy can yield a plethora of information about the local micro- and nano-scopic environment surrounding the paramagnetic impurity site.

To enhance the signal obtained from such impurity spin ensembles, and thereby reduce the measurement time, small-mode-volume and high-quality-factor resonators can be used~\cite{narkowicz_scaling_2008,shtirberg_high-sensitivity_2011,malissa_superconducting_2013,sigillito_fast_2014,artzi_induction-detection_2015,sidabras_extending_2019}. Additionally, noise can be reduced by cooling down the sample to temperatures much lower than $\hbar \omega_0/k_B$ and by using low-noise microwave amplifiers. However, vacuum fluctuations in the detection mode ultimately impose an upper limit to the achievable signal-to-noise ratio (SNR)~\cite{bienfait_reaching_2016,eichler_electron_2017,probst_inductive-detection_2017,ranjan_electron_2020}. Recently, an alternative detection method that potentially surmounts this limit was proposed and demonstrated~\cite{albertinale_detecting_2021}. It relies on detecting with a Single Microwave Photon Detector (SMPD)~\cite{romero_microwave_2009,helmer_quantum_2009,chen_microwave_2011,koshino_implementation_2013,sathyamoorthy_quantum_2014,kyriienko_continuous-wave_2016,inomata_single_2016,narla_robust_2016,besse_single-shot_2018,kono_quantum_2018,lescanne_irreversible_2020} the incoherent microwave photons spontaneously emitted by the spins when they relax radiatively after excitation by a $\pi$ pulse through the microwave resonator. This microwave fluorescence signal yields information about the spin density and lineshape as a Hahn echo, but with a larger signal and less noise ~\cite{albertinale_detecting_2021}. The spin fluorescence detection method was first applied to shallow-implanted bismuth donors in silicon, in a narrow range of magnetic fields~\cite{albertinale_detecting_2021}. It was more recently used for single-spin detection of erbium impurities in a $\mathrm{CaWO}_4$ crystal~\cite{wang_single-electron_2023}. The objective of this article is to demonstrate the broad applicability of this method for characterizing ensembles of paramagnetic impurities, to provide a detailed study of the fluorescence curves decay and their signal-to-noise ratios, and to report an example where the high sensitivity of this method yields new spectroscopic results difficult to obtain by other means.

In this study, we consider a scheelite crystal of calcium tungstate ($\mathrm{CaWO}_4$), on which a planar superconducting micro-resonator was deposited~\cite{le_dantec_twenty-three-millisecond_nodate} (see Fig.~\ref{fig1}). We record fluorescence-detected (FD) EPR spectra over a wide ($\sim 50$\,mT) range of magnetic fields in the crystal (a,b) plane, which shows the presence of several paramagnetic impurities. Because these impurities are distributed homogeneously throughout the crystal, while their radiative relaxation rate $\Gamma_R$ depends on their location within the microwave resonator mode, the time traces of the average SMPD count rate following an excitation pulse are non-exponential, and depend on the excitation pulse power. Focusing on the $\mathrm{Er}^{3+}$ $I=0$ transition as a model system~\cite{antipin_a._paramagnetic_1968,bertaina_rare-earth_2007,le_dantec_twenty-three-millisecond_nodate}, we reach quantitative agreement between the fluorescence measurements and a model in which the contribution of each spin-impurity is added incoherently. We then compare the SNR reached using ID- and FD-EPR under identical conditions, and find that it is larger with FD-EPR, thus confirming the results obtained in reference~\cite{albertinale_detecting_2021}. We finally focus on the spectra measured at very low excitation powers, which mostly probe spin-impurities close to the resonator inductance. We find that they are strongly distorted compared to those acquired at higher excitation power, which we interpret as being caused by strain shifts due to the differential thermal contraction of the niobium thin-film compared to the underlying crystal, as already observed for donors in silicon~\cite{pla_strain-induced_2018,ranjan_spatially-resolved_2021}. This allowed us to draw qualitative conclusions on the influence of strain on the gyromagnetic tensor of $\mathrm{Er}^{3+}:\mathrm{CaWO}_4$. Coherent Rabi oscillations are observed for the most highly strained impurity-spins, due to their localization in a region of space where the microwave field generated by the resonator is homogeneous. 

The article is organized as follows. Section II provides the necessary background on the spin-transition properties of paramagnetic ions in $\mathrm{CaWO}_4$ and on the modeling of the spin fluorescence curves. The experimental setup and sample are described in Section III. Section IV reports the FD spectroscopy measurements in the crystal (a,b) plane. The power-dependent fluorescence curves and their modeling, as well as the SNR discussion, are presented in Section V. Section VI is devoted to the observation of strain-induced lineshape distorsions of the erbium electron-spin transition, and of coherent oscillations in the highly-strained region of the spectrum. Conclusions and perspectives follow, in Section VII.

\section{Theoretical background}

\subsection{Radiative relaxation of an ensemble of paramagnetic impurities coupled to a resonator}

\subsubsection{Single-spin coupling Hamiltonian and dynamics}

We first consider one paramagnetic impurity coupled to a resonator of frequency $\omega_0$ by the magnetic dipolar interaction. We suppose that one of the spin system transitions has its frequency $\omega_s$ close to $\omega_0$, with the ground state noted as $|\downarrow \rangle$ and excited state as $|\uparrow \rangle$. We consider here that the temperature $T$ of the experiment satisfies $T \ll \hbar \omega_s / k$, so that the spin is in state $|\downarrow \rangle$ at thermal equilibrium. We can then model the spin system as an effective two-level system in the $|\downarrow \rangle,|\uparrow \rangle$ basis, yielding a Jaynes-Cummings-type Hamiltonian~\cite{haikka_proposal_2017}

\begin{equation}
H = \omega_0 a^\dagger a - \frac{\omega_s}{2} \sigma_z + g_0 (\sigma_+ a + \sigma_- a^\dagger),
\end{equation}

\noindent for the spin-resonator interaction, where $a$ ($a^\dagger$) is the annihilation (creation) operator for the resonator field, $\sigma_{z,+,-}$ the Pauli operator, and $g_0 = \langle \downarrow | \mathbf{S} | \uparrow \rangle \cdot \mathbf{\gamma} \cdot \mathbf{\delta B_1}$ the spin-resonator coupling constant, with $\mathbf{S}$ the electronic spin operator, $\mathbf{\gamma}$ the gyromagnetic tensor, and $\mathbf{\delta B_1}(\mathbf{r})$ the standard-deviation of the microwave magnetic field $\mathbf{B_1}$ at the spin location $\mathbf{r}$ when no photon is in the resonator. 

\begin{figure*}[tbh!]
    \includegraphics[width=0.8\textwidth]{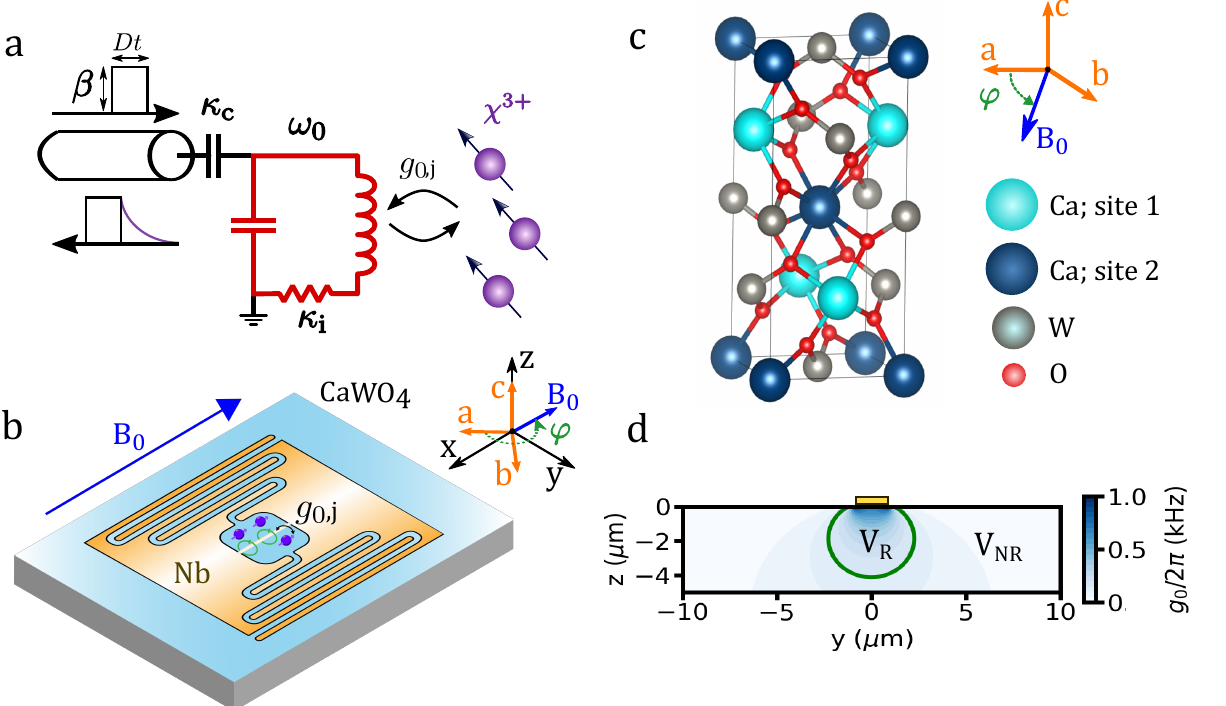}
    \caption{\label{fig1}
    \textbf{Experiment description.}
    \textbf{a}, schematic of the experiment. The resonator of frequency $\omega_0/2\pi$ is magnetically coupled to an ensemble of electron spins of trivalent ions ($\chi^{3+}$). The coupling strength to spin $j$ is $g_{0,j}$. The resonator energy is damped by coupling to the transmission line at a rate $\kappa_\text{c}$ and by internal losses at a rate $\kappa_{\text{i}}$. The spins are excited by microwave pulses of amplitude $\beta$ and duration $\Delta t$.
    \textbf{b}, sketch of the niobium planar micro-resonator fabricated on top of the $\mathrm{CaWO}_4$ sample in the $(a,b)$ plane. The $630 \, \mu \mathrm{m}$-long and $2\ \mu m$-wide wire is directed along the x-axis, making an angle $\varphi_\text{w}=51^\circ$ with the $a$-axis of the crystal, and $B_0$ is in the $ab$ plane, making an angle $\varphi$ with the $a$-axis of the crystal.
    \textbf{c}, schematic of the $\mathrm{CaWO_4}$ unit-cell, illustrating its tetragonal symmetry around the crystal $c$-axis. The calcium ions, which are the sites of insertion of the trivalent metallic ions, are located in two non-equivalent sites, shown in light and dark blue respectively, related by inversion symmetry. 
    \textbf{d}, cross section of the sample below the wire (shown as a gold rectangle) that displays the spatial distribution of the coupling $g_0$, which we assume to be constant along the wire length. The green contour marks the $\Gamma_{R} = \Gamma_{NR} = 0.15 s^{-1}$ limit between the Purcell volume ($V_R$, spins located inside the green contour) and the spin-lattice volume ($V_{NR}$, spins located outside the contour). 
    }
\end{figure*}

As shown in Fig.~\ref{fig1}, the resonator is capacitively coupled to a measurement line, so that the intra-resonator microwave energy is out-coupled at a rate $\kappa_c$. It also has internal losses, parametrized using the rate $\kappa_i$, so that the energy loss occurs at a total rate $\kappa = \kappa_i + \kappa_c$. Whereas spin-resonator dynamics in the Jaynes-Cummings Hamiltonian can be quite complex, here we will restrict ourselves to the so-called weak coupling regime $g_0 \ll \kappa$ (which is relevant to our experiments, as shown below), in which quantum correlations between the spin and the resonator can be neglected. Rectangular-shaped microwave pulses of amplitude $\beta$ and duration $Dt$ are sent to the resonator input. . They generate an intra-resonator field $\alpha(t) \mathrm{e}^{-i \omega_0 t}$, with $\alpha(t)$ given by the equation $\dot{\alpha}(t) = -\frac{\kappa}{2} \alpha + \sqrt{\kappa_c} \beta$. This field drives the spins with a Rabi frequency $2 g_0 \alpha(t)$; the subsequent spin dynamics can then be computed numerically using the Bloch equations~\cite{ranjan_pulsed_2020}.  

The coupling to the resonator opens a new relaxation channel for the spin through spontaneous emission of a microwave photon into the resonator, because of the Purcell effect~\cite{purcell_spontaneous_1946}. The corresponding radiative rate is given by: 

\begin{equation}\label{eq:GammaR}
\Gamma_R = \frac{4 g_0^2}{\kappa} \frac{1}{1 + (2 \delta / \kappa)^2 },    
\end{equation}

\noindent where $\delta = \omega_0 - \omega_s$ is the spin-resonator detuning~\cite{bienfait_controlling_2016}. This phenomenon is essential for the experiments reported here, since the fluorescence signal consists exactly of these emitted photons. The total spin relaxation rate $\Gamma_1 = \Gamma_R + \Gamma_{NR}$ also includes a non-radiative contribution $\Gamma_{NR}$, from the spin-lattice interaction. Note that $\Gamma_R$ strongly depends on the spin location $\mathbf{r}$ within the resonator mode (via the coupling constant $g_0$) and on its detuning with the resonator $\delta$, whereas the dependence of $\Gamma_{NR}$ on $\delta$ can essentially be neglected on the scale of the detection resonator linewidth $\kappa_c/2\pi$, which is on the order of $1$\,MHz. Consider then the spin prepared at $t=0$ in state $|\uparrow \rangle$. From standard input-output theory, it is straightforward to show that, for $t>0$, the probability of photon emission into the measurement waveguide per unit time is given by 

\begin{equation}
   |\alpha_{out}|^2 (t) =  \Gamma_R \frac{\kappa_c}{\kappa} \mathrm{e}^{- \Gamma_1 t}.   
\end{equation}

The total number of photons emitted in the waveguide is $\int |\alpha_{out}|^2 (t) dt = \frac{\kappa_c}{\kappa} \frac{\Gamma_R}{\Gamma_1}$, which approaches $1$ when the resonator internal losses and non-radiative spin relaxation rate are both negligible.

\subsubsection{Spin ensemble : fluorescence and echo}

In the experiments reported here, we will consider ensembles of $N$ paramagnetic impurities, distributed homogeneously throughout the crystal. Their electronic Larmor frequencies vary from spin to spin, due to slight changes in their local electrostatic or magnetic environment. This inhomogeneous broadening of the spin resonance is characterized by a detuning distribution $\rho(\delta)$ around the central frequency $\omega_s$ with standard deviation $\Gamma$, the so-called inhomogeneous linewidth.

The spin-resonator coupling constant $g_0$ is also inhomogeneously distributed, due to its dependence on the spin position within the resonator mode. Each spin $j$ has thus its own coupling constant $g_{0,j}$; in a continuous modeling, the coupling constant $\rho(g_0)$ distribution is introduced. An example of $g_0$ spatial distribution is shown in Fig.~\ref{fig1}, computed for the resonator design that is used in our experiment. $g_0$ achieves a maximum value of $g_0/2\pi \sim 1.0 $\,kHz close to the resonator inductance for our geometry, and gradually decreases for spins at further distances that are more weakly coupled with an approximate $\sim 1/r$ scaling. 

The $g_0$ distribution has two direct important consequences: the Rabi rotation angle $\int 2 g_0 \alpha(t) dt$ undergone by the spins under the application of a microwave pulse is itself inhomogeneously distributed, as well as the spin radiative rate $\Gamma_R$ (see Eq.~\ref{eq:GammaR}). For reasons that will be clarified later, it is useful to define a Purcell-volume $V_R$, in which spins that satisfy the condition $\Gamma_R (\delta = 0) > \Gamma_{NR}$ are located, and which is physically centered around the inductance, and the complementary volume $V_{NR}$ of weakly coupled spins (see Fig.~\ref{fig1}).

The coupling of $N$ spins to a resonator may give rise to collective enhancement or attenuation of radiation emission (ie: super- or sub-fluorescence), due to the emergence of quantum coherence in the ensemble via their coupling to a common mode~\cite{dicke_coherence_1954,temnov_superradiance_2005,lei_many-body_2023,solomon_anomalous_2023}. Here, we restrict ourselves to the limit in which the emission rate $N \Gamma_R$ of a photon by the spin ensemble is much smaller than the rate $\Gamma$ at which coherence is lost in the ensemble, that is, $N \Gamma_R \ll \Gamma$ (assuming for simplicity identical coupling constants $g_0$ for all $N$ spins and therefore identical radiation rates $\Gamma_R$). This weak-ensemble-coupling condition is more restrictive than the single-spin weak coupling condition. It corresponds to an ensemble
cooperativity $C \equiv 4 \frac{N g_0^2}{\kappa \Gamma}$ much smaller than $1$. In magnetic resonance terms, it is equivalent to stating that radiation damping is negligible. In this regime, the back-action of the field radiated by the ensemble has a negligible impact on the dynamics of each spin. The latter can therefore be computed independently for each spin, and the radiated field is simply the sum of the independent spin contributions.

\paragraph{Hahn echo Inductive Detection} As already mentioned in the introduction, the orthodox method of studying inhomogeneously broadened spin-ensembles in conventional EPR spectroscopy is through the measurement of Hahn echoes. The sequence consists of a $\pi/2_X$ pulse applied at $t=0$ at the resonator frequency $\omega_0$, generating immediately a transient transverse magnetization along $Y$ that quickly decays due to inhomogeneous broadening. After a waiting time $\tau$, a $\pi_Y$ pulse is applied, which refocuses the spin evolution and transiently restores the transverse magnetization $\langle S_Y \rangle$ after a second delay $\tau$, which is called a spin echo. 

Echoes are usually detected through the coherent microwave pulse inductively emitted along the $X$ quadrature in the detection waveguide by the Larmor precession of the magnetization. A simple formula for the SNR reached in this inductive detection can be derived, assuming ideal pulses, identical coupling constant $g_0$ for all spins, and a signal power loss $\eta$ (with $\eta = \kappa_C / \kappa$ if the resonator internal losses are the only ones present). Under these hypotheses, the amplitude of this spin echo pulse can be shown to be $\langle X_e \rangle = N \sqrt{\eta \Gamma_R / 2\Gamma}$ if $\Gamma \ll \kappa$, or $\langle X_e \rangle \sim N \sqrt{\eta \Gamma_R / 2\kappa}$ in the more common case $\Gamma \gg \kappa$~\cite{bienfait_reaching_2016}. Quantum mechanics imposes a minimum amount of noise to quadrature detection, with standard deviation $\delta X_e \geq 1/2$. The lower bound $\delta X_e = 1/2$ is approached experimentally by using amplifiers that reach the quantum limit~\cite{bienfait_reaching_2016,eichler_electron_2017,probst_inductive-detection_2017}, such as Josephson Parametric Amplifiers (JPAs). We obtain a maximum signal-to-noise ratio $\langle X_e \rangle / \delta X_e = N \sqrt{2 \eta \Gamma_R / \kappa}$ (assuming that $\Gamma \gg \kappa$). Even for an ideal experiment where $\eta = 1$, the SNR has an upper bound, due to quantum fluctuations~\cite{bienfait_magnetic_2017}, the highest figure demonstrated so far being $12\,\mathrm{spin}/\sqrt{\mathrm{Hz}}$~\cite{ranjan_electron_2020}. 

\paragraph{Fluorescence Detection}

The fluorescence detection method consists in counting the photons emitted by the spin ensemble after the application of a microwave pulse of amplitude $\beta$ at time $t=0$. The fluorescence signal is obtained by summing incoherently the contributions of each spin. Denoting $\langle S_{Z,j} \rangle$ as the average longitudinal magnetization of spin $j$ following the pulse, the number of photons emitted per unit time is

\begin{equation}
\label{eq:fluo}
    |\alpha_{out}|^2 (t) = \sum_j \eta \Gamma_{R,j} \frac{1 + 2\langle S_{Z,j}\rangle}{2} \exp^{- \Gamma_{1,j} t}.
\end{equation}

\noindent 
In the simplified model where inhomogeneities in the coupling constant and detuning are neglected and the exciting pulse is a perfect $\pi$ pulse, this equation becomes $|\alpha_{out}|^2 (t) = N  \Gamma_R \eta \mathrm{e}^{- \Gamma_1 t}$, and the total number of emitted photons is simply $\langle C \rangle =  N \eta \frac{\Gamma_R}{\Gamma_1}$. In this detection process, noise arises from the partition noise of the emitted photons, and from the SMPD noise, characterized by a background count rate $\alpha$ due to false positives (i.e., dark counts) and to input photons not emitted by the spins. The total noise is thus characterized by a standard deviation $\delta C = \sqrt{\alpha/\Gamma_1  + \eta (1-\eta) N }$. Therefore, the one-shot signal-to-noise ratio is expected to be $\langle C \rangle / \delta C = \eta N / \sqrt{\alpha / \Gamma_R + \eta (1-\eta) N}$, assuming for simplicity that only spins within $V_R$ are excited by the pulse. In an ideal experiment where $\eta = 1$ and $\alpha = 0$, this SNR has no upper bound and, therefore, can be larger than in ID EPR spectroscopy. This is due to the fact that an energy-sensitive detector such as the SMPD is not sensitive to the vacuum fluctuations of the microwave field. 

Taking into account the inhomogeneous distribution of coupling rates has a strong impact on the time dependence of the fluorescence signal, which we discuss qualitatively now and more quantitatively in Section V. This is easier to see when the contributions of spins in the Purcell volume $V_R$ and in the spin-lattice volume $V_{NR}$ are separated. Spins within $V_{NR}$ relax at approximately the same rate $\Gamma_1 \simeq \Gamma_{NR}$. Within the volume $V_R$ on the other hand, the relaxation rate $\Gamma_1 \simeq \Gamma_R$ varies from spin to spin. Equation \ref{eq:fluo} describes a signal that consists of a fast-relaxing, non-exponential component due to the spins within $V_R$, followed by an exponential tail decaying at the rate $\Gamma_{NR}$ that arises from spins within $V_{NR}$. Fluorescence-Detected EPR spectroscopy, therefore, allows one to observe the spin-lattice relaxation, simply by considering the exponential decay of the fluorescence curve at long time delays. This represents a valuable measurement advantage over ID EPR spectroscopy. Note also, that even though spins in $V_{NR}$ have a small probability $\frac{1 + 2\langle S_{Z,j}\rangle}{2} \Gamma_{R,j}/(\Gamma_{R,j} + \Gamma_{NR})$ of emitting one photon, their total contribution can be significant since there are many of them. Subsequently, the shape of the curve depends greatly on the excitation pulse amplitude. For low amplitudes, only the spins with the largest coupling constant (those within $V_R$) are significantly excited, and they will relax with a large Purcell rate. The overall decay rate reduces at higher excitation pulse amplitude, due to the increasing contribution of more weakly coupled spins.  

\subsubsection{Simulation of fluorescence curves}

Here we develop a simulation tool that computes Eq.~\ref{eq:fluo} for a given distribution $\rho(g_0)$ and $\rho(\delta)$, pulse parameters (amplitude $\beta$ and duration $Dt$), and repetition time $T_{rep}$ (defined as the time between the last pulse of a sequence and the first pulse of the subsequent one). 

For this purpose, we discretize $\rho(g_0)$ into bins of size $\Delta g$, and $\rho(\delta)$ in bins of size $\Delta \delta$. We can then simulate the evolution of a spin packet with coupling $g_{0,i}$ and detuning $\delta_j$ using the Bloch equations (including the Purcell and spin-lattice relaxation). The density matrix is initialized as follows: we consider that the previous sequence leaves all spins saturated (i.e., with $\langle S_{z,ij} \rangle = 0$). The spin packet then relaxes towards its equilibrium $\langle S_{z,ij} \rangle = -1/2$ with rate $\Gamma_1$, and we therefore take $\langle S_{z,ij} \rangle = -1/2 + 1/2 \mathrm{e}^{-[\Gamma_{NR} + \Gamma_R(g_{0,i},\delta_j)] T_{rep}}$ as the initial condition before the excitation pulse. Each simulation yields the spin packet longitudinal component $\langle S_{z,ij} \rangle$ at the end of the pulse sequence. We then compute Eq.~\ref{eq:fluo} by adding the contribution of each packet, with weight $\rho(g_{0,i}) \rho(\delta_j) \Delta g \Delta \delta$. 

\begin{figure}[t!]
    \includegraphics[width=\columnwidth]{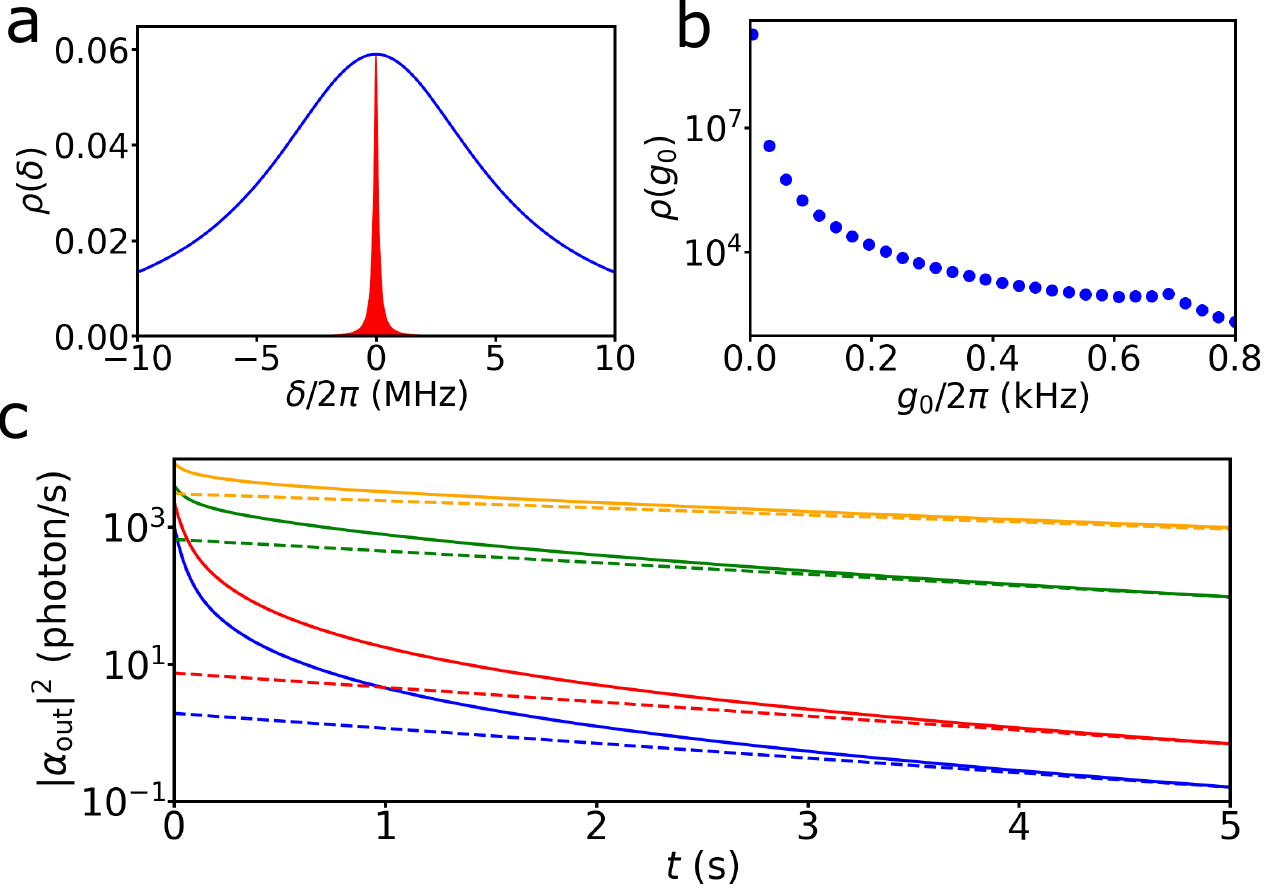}
    \caption{\label{fig2}
    \textbf{Spin simulation}.
    \textbf{a}, in blue, normalized frequency distribution $\rho(\delta)$ deduced from the measured $\text{Er}^{3+}$ linewidth in the conditions described in Section V. In red, resonator linewidth.
    \textbf{b}, spin density distribution $\rho(g_0)$ deduced from the spatial dependence of the coupling constant. The structure at $g_0/2\pi \approx 0.7$\,kHz is due to the wire width.
    \textbf{c}, solid lines are simulated spin fluorescence curves at various spin excitation powers. The decay is non-exponential at short times, due to the distribution of radiative decay rates $\Gamma_R$ in the Purcell volume $V_R$. The tail at long delay times is exponential, and corresponds to the contribution of spins in the spin-lattice volume $V_{NR}$. An exponential fit of the second half of each curve is shown as a dashed line. 
    }
\end{figure}

An example is shown in Fig.~\ref{fig2}. The distributions $\rho(g_0)$ and $\rho(\delta)$ correspond to the experimental parameters as described in Section III. Example fluorescence curves are shown in panel $c$, for various values of the pulse amplitude $\beta$. The curves display the characteristic features anticipated previously. Their decay at long delay times is exponential, with a rate $\Gamma_{NR}$, due to the contribution of spins in the spin-lattice volume $V_{NR}$. The faster decay at short delay is non-exponential, and arises from the distribution of radiative relaxation rates throughout spins in the Purcell volume $V_R$. We also confirm that the shape of the curve strongly depends on the excitation pulse amplitude $\beta$, because of the correlation between the Rabi frequency and the radiative relaxation rate introduced by the Purcell effect. 

\subsection{Paramagnetic impurities in a scheelite crystal}

\subsubsection{Crystal description and rare-earth spin Hamiltonian}

The scheelite crystal of $\mathrm{CaWO}_4$ used in this work is in the tetragonal crystal system, corresponding to the $I4_1/a$ space group. Its unit cell is shown schematically in Fig.~\ref{fig1}d. 

Paramagnetic impurities in $\mathrm{CaWO}_4$ can be found in a large variety of sites, but in this work we will encounter exclusively species that enter as substitutes for the $\mathrm{Ca}^{2+}$ ions. They can be found in two inversion-related unequivalent sites, both of them displaying $S_4$ symmetry (see Fig.~\ref{fig1}d). For metallic ions in the trivalent oxidation state (such as rare-earth-ions or REIs), charge compensation is necessary for their insertion into $\mathrm{CaWO}_4$ in substitution of $\mathrm{Ca}^{2+}$. Charge compensation seems to occur over large distance for most REIs, but not for $\mathrm{Fe}^{3+}$ as discussed in more details below. 

Ions with an odd number of electrons have a ground state that is at least twice degenerate in the absence of a magnetic field, due to time-reversal symmetry; these Kramers ions generally give the strongest EPR signals. Kramers doublets can be treated as an effective electronic spin $S=1/2$, with an associated spin Hamiltonian 

\begin{equation}
    H = \mathbf{B_0} \cdot \mathbf{\gamma} \cdot \mathbf{S} + \mathbf{I} \cdot \mathbf{A} \cdot \mathbf{S} + \gamma_n \mathbf{B_0} \cdot \mathbf{I},
\end{equation}

\noindent where $\mathbf{S}$ represents the electron spin operator, $\mathbf{I}$ the spin operator of the nuclear spin of the ion if the latter has one, $\mathbf{\gamma}$ the electron spin gyromagnetic tensor, $\gamma_n$ the nuclear spin gyromagnetic ratio, and $\mathbf{A}$ the tensor describing the hyperfine interaction between electron and nuclear spins. 

Owing to the $S_4$ symmetry of the $\mathrm{Ca}^{2+}$ sites, the gyromagnetic tensor is diagonal along the $(a,b,c)$ crystalline axes, with $\gamma_c \equiv \gamma_\parallel$ and $\gamma_a = \gamma_b \equiv \gamma_\perp$. For the high-spin trivalent iron $\mathrm{Fe}^{3+}$ on the other hand, which is believed to also substitute for calcium, charge compensation seems to occur predominantly in the immediate vicinity of the ion, probably through a calcium vacancy or another monovalent ion. As a result, the spectrum of $\mathrm{Fe}^{3+}:\mathrm{CaWO}_4$ consists of $4$ rotation-related lines, due to the $4$ different possible orientations of the next neighbor sites~\cite{kedzie_paramagnetic_1965,mcgavin_epr_1985,claridge_x-band_1997}. 

Relevant to this work are the gyromagnetic tensor elements for the ground-state doublets of erbium $\mathrm{Er}^{3+}$, ytterbium $\mathrm{Yb}^{3+}$, 
neodymium $\mathrm{Nd}^{3+}$ and iron $\mathrm{Fe}^{3+}$. They are summarized in Table \ref{tab1} for the $I=0$ isotopes of $\mathrm{Er}^{3+}$, $\mathrm{Yb}^{3+}$ and $\mathrm{Nd}^{3+}$, the $^{167}\mathrm{Er}^{3+}$ isotope which has nuclear spin $I=7/2$, the $^{171}\mathrm{Yb}^{3+}$ isotope with $I=1/2$, the $^{173}\mathrm{Yb}^{3+}$ isotope with $I=5/2$, the $^{143}\mathrm{Nd}^{3+}$ isotope with $I=7/2$, and the $^{145}\mathrm{Nd}^{3+}$ isotope with $I=7/2$. The parameters of the $^{56}\mathrm{Fe}^{3+}$ ground state doublet~\cite{claridge_x-band_1997} are given in the Appendix A for one of the $4$ rotation-related sites.

In addition to the hyperfine self-interaction $\mathbf{I} \cdot \mathbf{A} \cdot \mathbf{S}$, paramagnetic ions also interact with other nuclear spins in their magnetic environment, often referred to as the superhyperfine interaction. In $\mathrm{CaWO}_4$, the nuclear-spin environment consists mainly of the $^{183} \mathrm{W}$ isotope of tungsten, which is present with $14\%$ of the total tungsten abundance in a natural material, and has a nuclear spin $I=1/2$ with low gyromagnetic ratio $\gamma_W / 2\pi = 1.8$\,MHz/T. Therefore, $\mathrm{CaWO}_4$ is a good host matrix for minimsing magnetic decoherence, owing to its particularly low nuclear magnetic moment density; this was recently demonstrated with measurements of electron-spin coherence up to $23$\,ms in $\mathrm{Er}^{3+}:\mathrm{CaWO}_4$~\cite{ferrenti_identifying_2020,kanai_generalized_2021,le_dantec_twenty-three-millisecond_nodate,rancic_electron-spin_2022-1}.

\begin{table*}[tbh!]
\centering
\begin{tabular}{|c | c | c | c | c | c | c| c|} 
 \hline
 Isotope & Nuc. spin & Nat. abun. & $\gamma_{\parallel}/2\pi$ (GHz/T) & $\gamma_{\perp}/2\pi$ (GHz/T) & $A_{\parallel}/h$ (MHz) & $A_{\perp}/h$ (MHz) & $\gamma_n/2\pi$ (MHz/T) \\ 
 \hline
 $\mathrm{Er}^{3+}$ ($I=0$) & $0$ & 0.77 & 17.45 & 117.3 & 0 & 0 & 0\\ 
 $^{167}\mathrm{Er}^{3+}$ & $\nicefrac{7}{2}$ & 0.23 &  &  & 130 & 873 & 1.23 \\
 \hline
 $\mathrm{Yb}^{3+}$ ($I=0$) & $0$ & 0.7 & 14.7 & 54.85 & 0 & 0 & 0 \\
 $^{171}\mathrm{Yb}^{3+}$ & $\nicefrac{1}{2}$ & 0.14 &  &  & 788 & 3082 & 7.52 \\
 $^{173}\mathrm{Yb}^{3+}$ & $\nicefrac{5}{2}$ & 0.16 &  &  & 216 & 851 & 1.97\\
 \hline
 $\mathrm{Nd}^{3+}$ ($I=0$) & $0$ & 0.8 & 28.49 & 35.42 & 0 & 0 & 0 \\
 $^{143}\mathrm{Nd}^{3+}$ & $\nicefrac{7}{2}$ & 0.12 &  &  & 606 & 773 & 2.32 \\
 $^{145}\mathrm{Nd}^{3+}$ & $\nicefrac{7}{2}$ & 0.083 &  &  & 376 & 480 & 1.42 \\
 \hline
 \end{tabular}
 \caption{\textbf{Effective spin-Hamiltonian parameters for Er$\boldsymbol{^{3+}}$ and Yb$\boldsymbol{^{3+}}$ in CaWO$\boldsymbol{_4}$.} Values for $\mathrm{Er}^{3+}$  are taken from ref \cite{bertaina_rare-earth_2007}, for $\mathrm{Yb}^{3+}$ from \cite{sattler_electron_1970} and for $\mathrm{Nd}^{3+}$ see Appendix A.}
 \label{tab1}
\end{table*}

\subsubsection{Effect of electric field and strain on the gyromagnetic tensor}

\paragraph{Electric field shifts}

Since the anisotropy of the gyromagnetic tensor $\mathrm{\gamma}$ results from the action of the crystal field on the ion energy levels, it is perhaps not surprising that $\mathrm{\gamma}$ would be modified if an additional electric field were externally applied. The perturbation is generally linear in field amplitude for sites without inversion symmetry, which is the case for the substitutional sites of $S_4$ symmetry relevant to REIs in $\mathrm{CaWO}_4$ (see Fig.~\ref{fig1}). This Linear Electric Field Effect~\cite{schweiger_principles_2001} was first observed by Mims for $\mathrm{Ce}^{3+}:\mathrm{CaWO}_4$~\cite{mims_electric_1964}, then further studied for four paramagnetic REIs including erbium and ytterbium~\cite{mims_electric_1965}, and modeled semi-quantitatively from crystal-field theory~\cite{kiel_theory_1966,kiel_electric-field-induced_1970}. Here, we will outline the main results that are necessary for interpreting the spectra recorded in this work. 

In all the experiments described here, the magnetic field $B_0$ is applied in the $(a,b)$ plane, making an angle $\varphi$ with respect to the $a$ axis. In that case, due to the $S_4$ symmetry, it has been shown~\cite{mims_electric_1965} that only electric field components parallel to the crystal c-axis can modulate the gyromagnetic tensor, to first order. Under a small perturbative field $dE_c$, rotational symmetry of the gyromagnetic tensor around the $c$ axis is lifted, and two orthogonal principal axes emerge, at directions that do not correspond to the crystalline axes $a,b$. Along these new axes, the value of the gyromagnetic ratio is changed by $\pm \delta \gamma = K dE_c$. Defining $\varphi_0$ as the angle made by one of these axes with the $a$ axis, the spin resonance frequency change can be written as $\delta \omega (\varphi) = K \sin (2 \varphi - 2 \varphi_0) dE_c B_0$~\cite{mims_electric_1965}. In particular, when $B_0$ is applied along the direction $\varphi_0$, no shift in the spin transition frequency is observed. The value of $\varphi_0$ depends on the REI, and is $31^{\circ}$ and $47^{\circ}$ for $\mathrm{Er}^{3+}$ and $\mathrm{Yb}^{3+}$, respectively~\cite{mims_electric_1965}. 

Applied electric fields have another remarkable effect on impurities in $\mathrm{CaWO}_4$, as noticed by Mims~\cite{mims_electric_1964}. This comes from the fact that the crystal hosts two sites related by inversion symmetry. Under an applied electric field, an ion located in one site will undergo a positive frequency shift $\delta \omega(\varphi)$, while an ion in the inverted site will experience the opposite (i.e.: negative) shift $- \delta \omega(\varphi)$. Since both sites are chemically identical and thus equally populated in the crystal, an electric field applied along $c$ will shift one half of the ions by $+ \delta \omega (\varphi)$ and the other hald by $- \delta \omega (\varphi)$. This was illustrated through a very elegant experiment by Mims~\cite{mims_electric_1964}, where an electric field was applied during the first half of an echo sequence, such that two spin packets with opposite phases were generated, whose emission interfered leading to complete echo suppression at time $\tau=(2k + 1) \pi / 2 |\delta \omega| $. Similar observations were reported in the optical domain~\cite{macfarlane_optical_2007} (and references therein).

In our experiments, however, no static electric field is purposely applied. Instead, inhomogeneous internal electric fields are present throughout the crystal, due to the random distribution of charge impurities and defects (among which, charge compensation of the $\mathrm{Ca}^{2+}$ substitution by a trivalent REI), leading to a $\varphi$-dependent broadening of the EPR line. In $\mathrm{CaWO}_4$, these inhomogeneous electric shifts are frequently the dominant cause of inhomogeneous broadening. A clear signature is the strong dependence of $\Gamma$ on the angle $\varphi$, as observed first by Mims~\cite{mims_broadening_1966} and more recently by Le Dantec et al.~\cite{le_dantec_twenty-three-millisecond_nodate}, with $\Gamma$ reaching a minimum when $\varphi = \varphi_0$. Note that broadening by internal electric fields necessarily leads to symmetrical lineshapes, since the inversion-related sites are shifted in opposite frequencies. 

\paragraph{Strain shifts}

While the effect of applied electric fields was studied in detail by Mims and collaborators, the effect of mechanical strain has not yet been considered, to the best of our knowledge, nor observed experimentally. As will be shown in the last section of this article, here we observe clear evidence of such strain shifts. Hence, we tentatively propose a qualitative approach to understand these data, while hoping that a more serious modeling of strain shifts shall be developed in the future.

Strain shifts can be discussed by comparison with electric field shifts since they both rely on a perturbative modification of the crystal field. Even though they share a common electrostatic origin, we expect a different behavior for both. On the one hand, the electric shifts arise from the combination of the internal odd crystal field terms and the externally applied field (with odd-parity as well). This statement has been exploited quantitatively by Kiel to explain the observations under applied electric field using the so-called equivalent even crystal field approach~\cite{kiel_electric-field-induced_1970}. As a direct consequence of the parity, inversion-related sites (site 1 and 2) exhibit opposite Stark shift.
On the other hand, strain shifts can be simply interpreted as a modification of the original even crystal-field terms under the application of pressure. In contrast to the previous case, inversion-related sites should exhibit the same strain shift. As an observable consequence, strain induced broadening can lead to a asymmetrical lineshapes depending on the strain profile, as opposed to the symmetrical lineshapes expected from the electric field broadening when combined for both sites.

Beyond this reasoning, which is based on parity, we question the existence of a minimal strain dependence in relation to the observation of the electric insensitivity when the magnetic field is aligned along the direction $\varphi_0$. It is indeed likely that the direction $\varphi_0$ also exhibits a minimal sensitivity to strain, somehow generalizing the zero first-order electric shift for this orientation. Mims originally observed a minimal sensitivity to electric field which can be composed of high order spatial modes (multipole expansion) for sufficient close charges in the vicinity of the erbium ion and not only a uniform field as applied by external electrodes (linear shift). Strain precisely couples to higher-order spatial modes, so we also expect an insensitivity to strain in the direction $\varphi_0$. Consequently, we first speculate a minimal strain broadening at this point corresponding to a cancellation of the sensitivity. Additionally, lineshape at $\varphi_0 - d\varphi$ should then be the mirror image of the one observed at $\varphi_0 + d\varphi$ if the strain sensitivity also changes sign at $\varphi_0$. These qualitative considerations are experimentally probed and confirmed in Section VII.

\section{Experimental methods}

\subsection{Sample}

The sample used in this study was described in~\cite{le_dantec_twenty-three-millisecond_nodate}, to which we refer the reader for further details. It originates from a $\mathrm{CaWO}_4$ crystal grown for dark-matter detection experiments, using high-purity natural-abundance materials~\cite{erb_growth_2013}. A $3$\,mm by $6$\,mm rectangular slab with $0.5$\,mm thickness parallel to the $c$ axis was prepared from it. The sample was characterized by inductively-detected EPR spectroscopy at $7$\,K and $10$\,mK. In particular, concentrations of the REIs $[\mathrm{Er}^{3+}]=0.7 \pm 0.1$\,ppb and $[\mathrm{Yb}^{3+}]= 38 \pm 4$\,ppb were measured. Other paramagnetic impurities were observed, which are confirmed in the following section. X-ray diffraction measurements indicate that the normal to the sample plane is not exactly parallel to the $c$ axis, but makes an angle $\theta_c = 3^{\circ}$ with it; this is taken into account in the spin spectra simulations presented in Section IV.

Three superconducting micro-resonators are patterned on top of this sample, one of which had a resonance frequency $\omega_0 / 2\pi  = 6.999 $\,GHz and was used exclusively for this work. It corresponds to `resonator 1' presented in reference~\cite{le_dantec_twenty-three-millisecond_nodate} and is also described in reference\cite{billaud_microwave_2023}. The central element of this resonator is a $630 \mu \mathrm{m}$-long, $2 \mu \mathrm{m}$-wide wire, which serves as an inductance for the resonator. The resonator impedance $Z_0 = 35 \Omega$ is estimated from microwave simulations, from which the vacuum current fluctuations $\delta i_0 = \omega_0 \sqrt{\hbar /2Z_0}$ are also obtained. By computing the spatial distribution of the current within the superconducting wire~\cite{bienfait_reaching_2016}, we obtain the spatial profile of the vacuum magnetic field around the wire $\mathbf{\delta B_1 (\mathbf{r})}$ and the spatial profile of the coupling constant $g_0(\mathbf{r})$ (see Fig.\ref{fig1}). We then further obtain the coupling constant distribution $\rho(g_0)$ (see Fig.\ref{fig2}).

\subsection{Experimental setup}

\begin{figure}[t!]
    \includegraphics[width=\columnwidth]{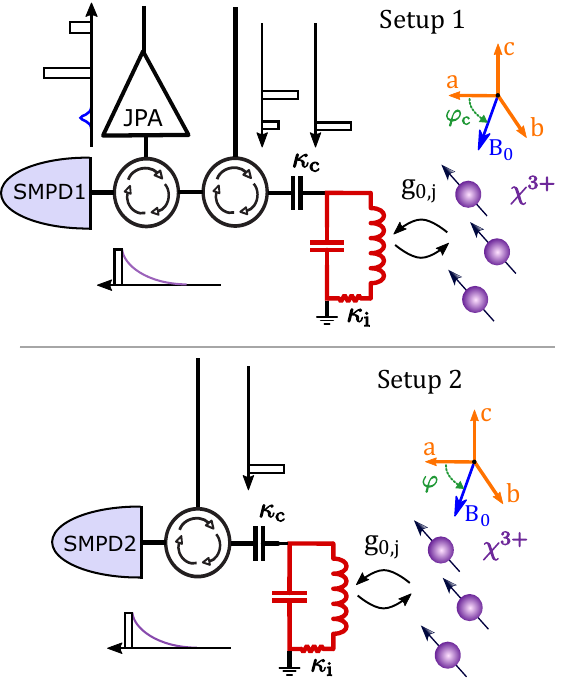}
    \caption{\label{fig3}
    \textbf{Schematic of the two low-temperature setups.} Circulators route the excitation pulses (black squares) towards the microwave resonator, while the emitted spin-fluorescence (red) and spin-echoes (blue) are routed to the microwave detectors. Black arrows indicate the direction of pulse and signal propogation. \textbf{Setup 1}. A JPA is installed behind the SMPD1 for ID echo measurement. $B_0$ is applied at a fixed angle $\varphi_c = 47^{\circ}$.
    \textbf{Setup 2}. SMPD2 is used, and $B_0$ is applied with a tunable angle $\varphi$.
    }
\end{figure}

As described in references~\cite{le_dantec_twenty-three-millisecond_nodate} and \cite{billaud_microwave_2023}, the sample is mounted in a copper sample-holder, and coupled to the measurement line through a plunger antenna, which allows for tuning of $\kappa_c$ by adjusting its length upon sample mounting. The results presented in this paper were recorded using two different setups, both of which are presented in Fig.~\ref{fig3} and are henceforth referred to as `setup 1' and `setup 2'. Reflection measurements recorded at $10$\,mK using setup 1 yield $\omega_0/2\pi = 7.004$ GHz, $\kappa_c = 8.2 \times 10^{5} \mathrm{s}^{-1}$ and $\kappa_i = 6.3 \times 10^{5} \mathrm{s}^{-1}$, whereas in setup 2 they yield $\omega_0/2\pi = 6.999$ GHz, $\kappa_c = 1.9 \times 10^{6} \mathrm{s}^{-1}$ and $\kappa_i = 3.6 \times 10^{5} \mathrm{s}^{-1}$. These numbers differ between setup 1 and setup 2, likely due to differences in the niobium film oxidation, in the out-of-plane static magnetic field and in the sample-holder shape and antenna to sample capacitance.

A static, homogeneous magnetic field $B_0$ is applied parallel to the sample. The angle $\varphi$ from the $a$ axis is set to $\varphi_c = 47^{\circ}$ in Setup $1$; it is tunable in Setup $2$ thanks to the use of $2$ orthogonal sets of home-made Helmholtz coils. 


Rectangular drive pulses of duration $Dt$ and peak power $P_{in}$ are applied to the sample through an input line with total attenuation $A$, corresponding to a drive amplitude at the sample input $\beta = \sqrt{P_{in}/(A \hbar \omega_0)} $. The attenuation is distributed along the low-temperature stages to minimize thermal radiation, with approximately $20$\,dB at 4K and $60$\,dB at 10 mK. We define a pulse strength $\epsilon = \beta \times Dt$ in order to compare different experiments in terms of Rabi angle that the spins undergo. A circulator is placed in front of the sample to route the drive pulses towards the resonator input, and the reflected pulses together with the spin emission towards an output line connected to the input of a SMPD. In Setup 1, a Josephson Parametric Amplifier is moreover connected behind the SMPD, enabling a quantitative comparison between measurements by photon counting and by quantum-limited quadrature detection; this is used in the experiments described in Section VI,.

The SMPD is based on a superconducting transmon qubit, and is operated under the same conditions described in ~\cite{lescanne_irreversible_2020} and ~\cite{albertinale_detecting_2021}. Two different SMPD devices are used in this work, characterized by different background count rates: SMPD1 described in reference ~\cite{albertinale_detecting_2021} with $\alpha_1 = 2 \cdot 10^{3} \mathrm{s}^{-1}$, and SMPD2 with $\alpha_2 = 5 \cdot 10^{2} \mathrm{s}^{-1}$. 

The SMPD is a narrow-band detector, and its center frequency needs to be tuned~\cite{albertinale_detecting_2021} at resonance with the spin resonator frequency $\omega_0$ in order to detect the spin fluorescence signal. Moreover, the resonator frequency depends weakly on $B_0$ due to the kinetic inductance contribution; therefore, the SMPD tuning needs to be regularly repeated during scans where $B_0$ varies over a large range, otherwise the fluorescence signal may not be detected if the detuning between the resonator and the SMPD center frequency is larger than its bandwidth. In that goal, an automatic procedure was designed. The procedure consists in measuring the resonator response with the SMPD, and tuning the SMPD center frequency to match it.

\subsection{Data acquisition}

We now present data acquisition methods for photon counting signals as well as quadrature measurements of the spin echoes.

\begin{figure*}[tbh!]
    \includegraphics[width=1\textwidth]{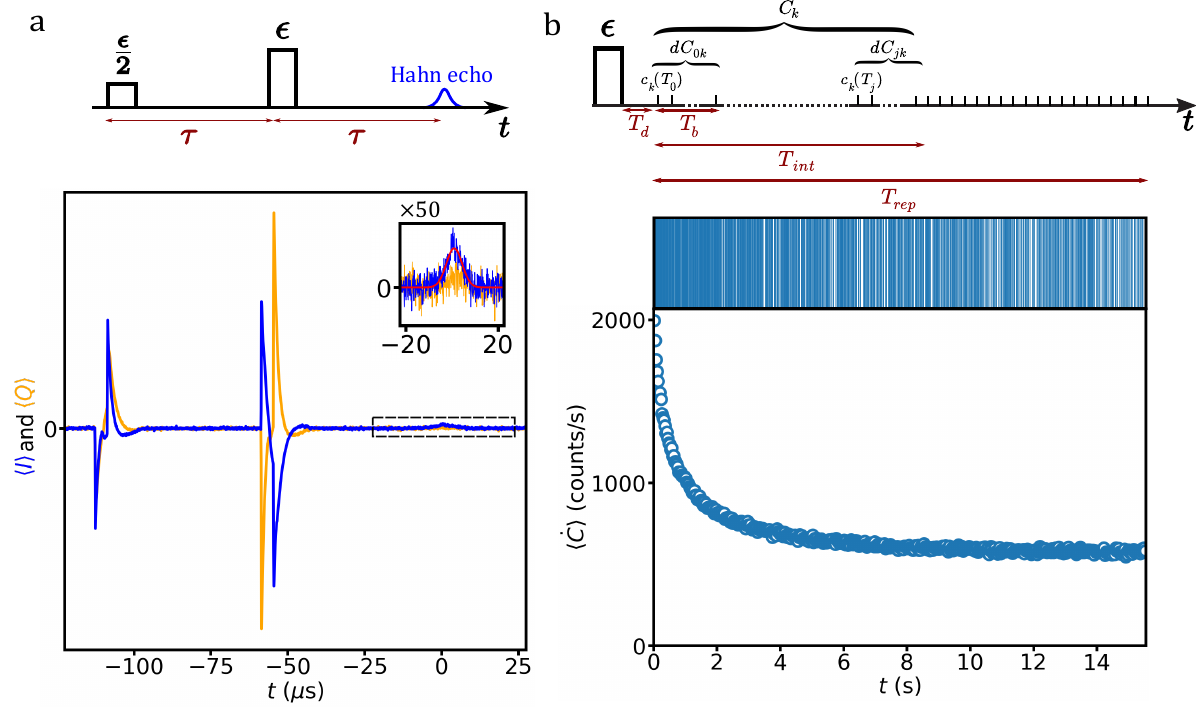}
    \caption{\label{fig4}
    \textbf{Typical data acquisition process for ID-EPR and FD-EPR measurements.} \textbf{a}, a, ID-EPR data acquisition. Top panel: schematic pulse and signal envelope. Bottom panel: Detected signal envelope, with the two quadratures (I in blue, Q in orange) plotted in arbitrary units, showing a Hahn echo at $t=0$. The inset in the top right shows an exploded view of the echo with a gaussian fit (red) yielding the echo envelope.
    \textbf{b}, FD-EPR sequence representation and typical experimental data. Top panel: manipulation sequence to detect spin fluorescence where we send at $t=0$ an excitation pulse of strength $\epsilon$ (black rectangle), wait a dead time $T_{\text{d}}$, then we cycle the SMPD over a time $T_{\text{rep}}$ (vertical bars). The SMPD clicks are treated (quantities on top) over various time (arrows at the bottom) as binned click probability or as integrated counts. Middle panel: SMPD clicks (vertical bars) following a spin excitation pulse, where we see a higher click probability just after the pulse. Bottom panel: averaged count rate $\langle \dot{C} \rangle$.
    }
\end{figure*}

A typical photon counting sequence is shown schematically in Fig.~\ref{fig4}b. $N_r$ identical excitation-measurement sequences are repeated. Each consists of an excitation pulse applied to the resonator input at $t=0$, followed by repeated SMPD acquisition cycles during a time $T_{rep}$. The presence or absence of a photon at the SMPD input during each cycle (of approximate duration $12 \mu$s) is mapped to the qubit state, which is then read-out, yielding time traces of binary digits $c(t_i)$, where $t_i$ is cycle $i$ time. The repetition time $T_{rep}$ is generally chosen so that $\langle \dot{C} \rangle (t)$ reaches approximately its steady-state value before the next excitation pulse is sent; this time depends on the excitation amplitude $\beta$ (defined previously) and is therefore adjusted according to $\beta$ to minimize the acquisition time (see Fig.~\ref{fig4} caption). The strong excitation microwave pulse perturbs the SMPD during a short time which we refer to as the dead-time $T_d$; here we take $T_d = 50 \mu \mathrm{s}$ (which is small compared to the typical sequence duration $\sim 1$\,s), during which the SMPD outcome is disregarded. 

A number of quantities can be extracted from the time traces $c_k(t_i)$, $1\leq k \leq N_r$ being the sequence index. First, the total number of counts during an integration time $T_{int}$ in sequence $k$ is given by $C_k = \sum_{T_d}^{T_{int}} c_k(t_i)$, from which we also obtain the ensemble-averaged number of counts $\langle C \rangle = \sum_{k=1}^{N_r} C_k / N_r$ as well as the standard deviation $\delta C = \sqrt{ \langle C^2 \rangle - \langle C \rangle^2 }$. Finally, it is important to determine the time dependence of the count rate after the pulse. For this, we choose a "coarse-grained" time window of size $T_b \ll T_{rep}$. The average count rate at time $T_j = T_d + j T_b $ is defined as $\langle \dot{C} \rangle (T_j) = \frac{1}{N_r}\sum_1^{N_r} dC_{jk} / T_b $, with $dC_{jk} = \sum_{[T_j,T_{j+1}]} c_{k}(t_j)$, which yields the fluorescence curves. 

To separate the counts due to spin relaxation from those attributed to background counts, we also define background-subtracted quantities. The simplest way to do so is to measure the dark count rate $\alpha$ in a separate controlled measurement, and to subtract this rate from the measured rate $\langle \dot{C} \rangle $. This approach, however, assumes that the background count rate is invariant under all measurement conditions. This is not the case, for instance the background count rate is found to fluctuate over the course of long data acquisitions. Moreover, the background count rate has been observed to depend on the microwave pulse power, possibly due to two-level-system heating, which therefore makes the subtraction measurement-specific. To minimize the impact of these variations, for certain datasets, we perform background subtraction by subtracting the count rate obtained at the end of the acquisition sequence (at time $T_{rep}$), which yields the spin contribution $\langle C_\mathrm{spin} \rangle = \langle C \rangle - T_{int} \langle \dot{C} (T_{rep}) \rangle$. Note that by doing so, we may also subtract some contribution from the spins that have not yet relaxed, if the repetition time is not much longer than the total spin relaxation time $\Gamma_1^{-1}$. This background-subtraction procedure may lead to some distortions, as will be seen in Section IV in particular. 

In addition to fluorescence, we also detect Hahn echoes, by microwave amplification followed by demodulation, following the usual inductive-detection (ID) approach. The sequence begins with a pulse of strength $\epsilon/2$, followed after a delay $\tau$ by a refocusing pulse of strength $\epsilon$. The echo amplitude $X_e$ is then obtained by integrating the signal over the echo temporal mode, which is determined by the ensemble-averaged echo time trace $\langle X_e \rangle (t)$. An example of such an averaged trace is shown in Fig.~\ref{fig4}a. The echo amplitude is converted into dimensionless units by using the SMPD as an absolute intensity detector ~\cite{billaud_microwave_2023}. Note that Fluorescence-Detection can also be used to detect echoes with larger SNR, as discussed in reference~\cite{billaud_microwave_2023}.

\section{Fluorescence-Detected Spectroscopy of Paramagnetic Impurities in Scheelite}

\begin{figure*}[tbh!]
    \includegraphics[width=1\textwidth]{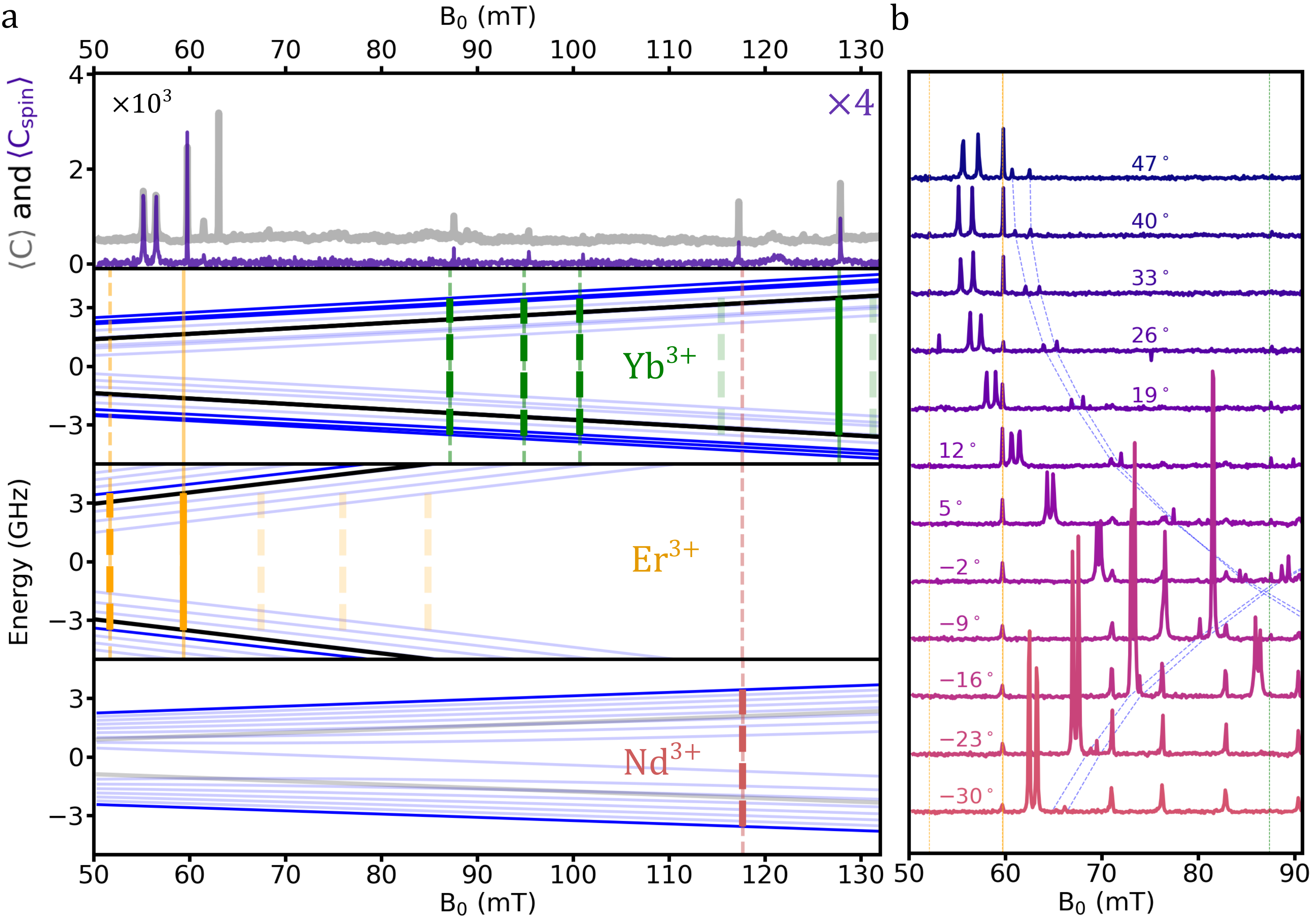}
    \caption{\label{fig5}
    \textbf{Spectroscopy and rotation pattern.} 
    \textbf{a}, shows $\langle C \rangle$ (gray) and $\langle C_{\mathrm{spin}} \rangle$ (purple) integrated over a time $T_{\text{int}} = 1.5$ s as a function of $B_0$, varying between 50.5 and 141 mT in steps of 0.05 mT, applied at an angle $\varphi = 37^\circ$. Background subtracted data has been multiplied by a factor 4 for a better comparison. The excitation pulse has an amplitude $\beta = 117 \, \text{ns}^{-1/2}$ and duration $\Delta t = 5 \, \mu s$, with a repetition time $T_{\text{rep}} = 2$ s. Every $2$\,mT, we tune the SMPD frequency and excitation pulse frequency to follow the resonator frequency shift caused by the magnetic field $B_0$.  
    \textbf{b}. From top to bottom : Blue solid lines are computed energy levels of $^{171}\text{Yb}^{3+}$, $^{173}\text{Yb}^{3+}$ (green), $^{167}\text{Er}^{3+}$ (orange) and $^{143}\text{Nd}^{3+}$ (red), as a function of $B_0$. Black solid lines are the energy levels of the corresponding $I=0$ isotopes. The magnetic field at which allowed electron-spin transition frequencies equal $\omega_0/2\pi$ are found are denoted as vertical bars, solid when $I = 0$ and dashed when $I\neq0$. The transparency of pairs of levels for which the transition was not observed has been increased for clarity.
    \textbf{c}, rotation pattern spectroscopy with background subtraction. $\langle C_{\text{spin}} \rangle$ is plotted as a function of $B_0$ and $\varphi$. The background is measured as the mean value of $\langle C \rangle$ in the time window $[4, 5]$ s. The angle of the magnetic field $\phi$ is displayed on top of each measurement. Pulse parameters are $T_{\text{int}} = 1.5$ s, $\beta = 58 \, \text{ns}^{-1/2}$, $\Delta t = 3 \, \mu s$, $T_{\text{rep}} = 5$ s and $B_0$ varying between 50.5 to 90.8 mT by step of 0.1 mT. Re-calibration of the SMPD is performed every 2 mT. Dashed blue lines show the expected magnetic field value at the resonance frequency of the $4$ orientations of the $\mathrm{Fe}^{3+}:\mathrm{CaWO}_4$ ground-state Kramers doublet.
    }
\end{figure*}

Fluorescence-Detected spectra were measured by recording the number of counts $\langle C \rangle$ detected after an excitation pulse, as a function of applied magnetic field $B_0$. The pulse amplitude  at the sample input $\beta = 117 \mathrm{ns}^{-1/2}$ was the highest that could be applied in our setup and its duration is $5 \mu s$. The repetition time was $T_{rep} = 2$s, and the number of counts is integrated over $T_{int} = 1.5$s. The resonator frequency shifts by 1.5 MHz between 50 and 130 mT due to the contribution of the kinetic inductance. As such, the SMPD automated tuning was done every $2$\,mT step, corresponding to a shift of $\approx 20$\,kHz, well within the SMPD and the pulse bandwidth. The data shown in this section are obtained using setup $2$.



The resulting spectra are shown in Fig.~\ref{fig5}a, for $\varphi = 37^{\circ}$. The step size is 0.05 mT, and the scan lasted 68 hours. Two spectra containing the same data are shown in Fig 5a; the difference between the two being the absence (gray) or presence (black) of background-subtraction, which is discussed in more detail in the following paragraphs. In both spectra several peaks are clearly visible; some of them can be identified by comparison to the energy levels of scheelite paramagnetic impurities whose gyromagnetic tensors are known. The $I=0$ $\mathrm{Er}^{3+}$ peak in particular is easily identified at $B_0 = 59.75$\,mT. This transition was used to precisely calibrate the field-to-current ratio of our Helmholtz coils by considering the resonator frequency ($6.999$\,GHz) and the known $\mathrm{Er}^{3+}$ gyromagnetic tensor. The first electron-spin transition of the $^{167}\mathrm{Er}^{3+}$ isotope within the spectral window is also observed, despite its low amplitude. The other seven were not observed, likely due to the thermal polarisation of the $^{167}\mathrm{Er}^{3+}$ hyperfine population at the experimental temperature of $10$\,mK. Ytterbium peaks are also observed in the spectrum; the $I=0$ transition shows a peak at exactly the expected $B_0 = 127.75$\,mT. The lowest energy level of $^{171}\mathrm{Yb}^{3+}$ is also visible, as well as the two lowest energy level transitions of $^{173}\mathrm{Yb}^{3+}$. The narrow line at $118$\,mT is attributed to the lowest energy-level transition of the $^{143}\mathrm{Nd}^{3+}$ isotope (see Appendix). The simulations used to identify the different species ignore the shift of the resonator frequency, since a shift of 1.5 MHz would approximately correspond to a shift of 0.03 mT for the $I=0$ Yb$^{3+}$, which is smaller than the  0.05 mT step-size of the spectra.

The same spectrum is shown with and without background subtraction in Fig.~\ref{fig5}a. We first see that the background fluctuations visible in the total count rate $\langle C \rangle (B_0)$ are indeed largely suppressed in the background subtracted $\langle C_{spin} \rangle (B_0)$. This makes it possible in particular to observe two unidentified broad preaks, one at $122$\,mT, and another at $128$\,mT, which are masked by background fluctuations in the spectrum without background subtraction. However, it is also clear that the amplitudes of most peaks are attenuated to a greater or lesser degree. This is particularly true for the two peaks at $61 - 62$\,mT, one of them being even completely suppressed. As will be clear below, this is due to the different timescales of the fluorescence curve decay on each peak. Because background subtraction uses the count rate at time $T_{rep}$, paramagnetic species for which the decay timescale is much longer than $T_{rep}$ exhibit peak amplitudes that are considerably suppressed relative to short-$T_1$ peaks. Background-subtracted spectra should therefore be used with care, in particular when it comes to obtaining quantitative information about the various paramagnetic impurity concentrations.

A background-corrected rotation pattern in the (a,b) plane, shown in Fig.~\ref{fig5}c, provides further information for the peak assignment. Note that in these spectra, the sharp peak that was completely suppressed by the background subtraction in Fig.~\ref{fig5}a is visible, due to a longer repetition time $T_{rep} = 5$\,s. As expected, the $\mathrm{Er}^{3+}$ and $\mathrm{Yb}^{3+}$ peak average position does not change due to their $S_4$ symmetry. Their width on the other hand is seen to depend appreciably on $\varphi$, as expected due to the effect of inhomogeneous electric fields discussed in Section II~\cite{le_dantec_twenty-three-millisecond_nodate}. Four other peaks whose rotation patterns are consistent with $S_4$ symmetry, visible only when $\varphi$ is lower than $15^\circ$, are also observed between $70$ and $90$\,mT. Their linewidths and amplitudes show a strong dependence on $\varphi$, and are identical amongst the four peaks, suggesting that they originate from the same paramagnetic impurity; we were unfortunately not able to match these transitions with any known paramagnetic impurity in $\mathrm{CaWO}_4$. Their linewidth minimum is around $\varphi = - 17^{\circ}$, and they are already strongly broadened once $\varphi = 15^{\circ}$, which suggests why they are not visible in the spectra taken at larger values of $\varphi$. 

In addition, two series of $4$ rotation-related peaks in non-$S_4$ symmetry are visible in the rotation pattern of Fig.~\ref{fig5}b. One series matches well with the expected rotation pattern for the ground-state Kramers doublet of $\mathrm{Fe}^{3+}$ in orthorombic symmetry (see Section II) reported in reference~\cite{claridge_x-band_1997}, as shown in Fig.~\ref{fig5}b. This series corresponds to the two peaks visible at $62-63$\,mT in Fig.~\ref{fig5}a, which were also reported in~\cite{le_dantec_twenty-three-millisecond_nodate} and labeled as $s1$ and $s2$. Note, however, that the quantitative agreement was obtained with the parameters of a Kramers doublet that was assigned to the high-energy doublet in~\cite{claridge_x-band_1997}; our data prove that this doublet is instead the ground state of $\mathrm{Fe}^{3+}:\mathrm{CaWO}_4$. Additional measurements and discussions regarding this series of peaks are presented in Appendix A. The other series of peaks (corresponding to the peaks labeled as $b1$ and $b2$ in reference~\cite{le_dantec_twenty-three-millisecond_nodate}) was still not identified here. All that was inferred is that these peaks likely originate from an $I=0$ species; that they seem to have a sizeable zero-field splitting (see ~\cite{le_dantec_twenty-three-millisecond_nodate}), and that their $T_1$ appears similar to erbium (see below). This may tentatively point to a non-Kramers REI species in a rhombic symmetry. 




\begin{figure}[t!]
    \includegraphics[width=\columnwidth]{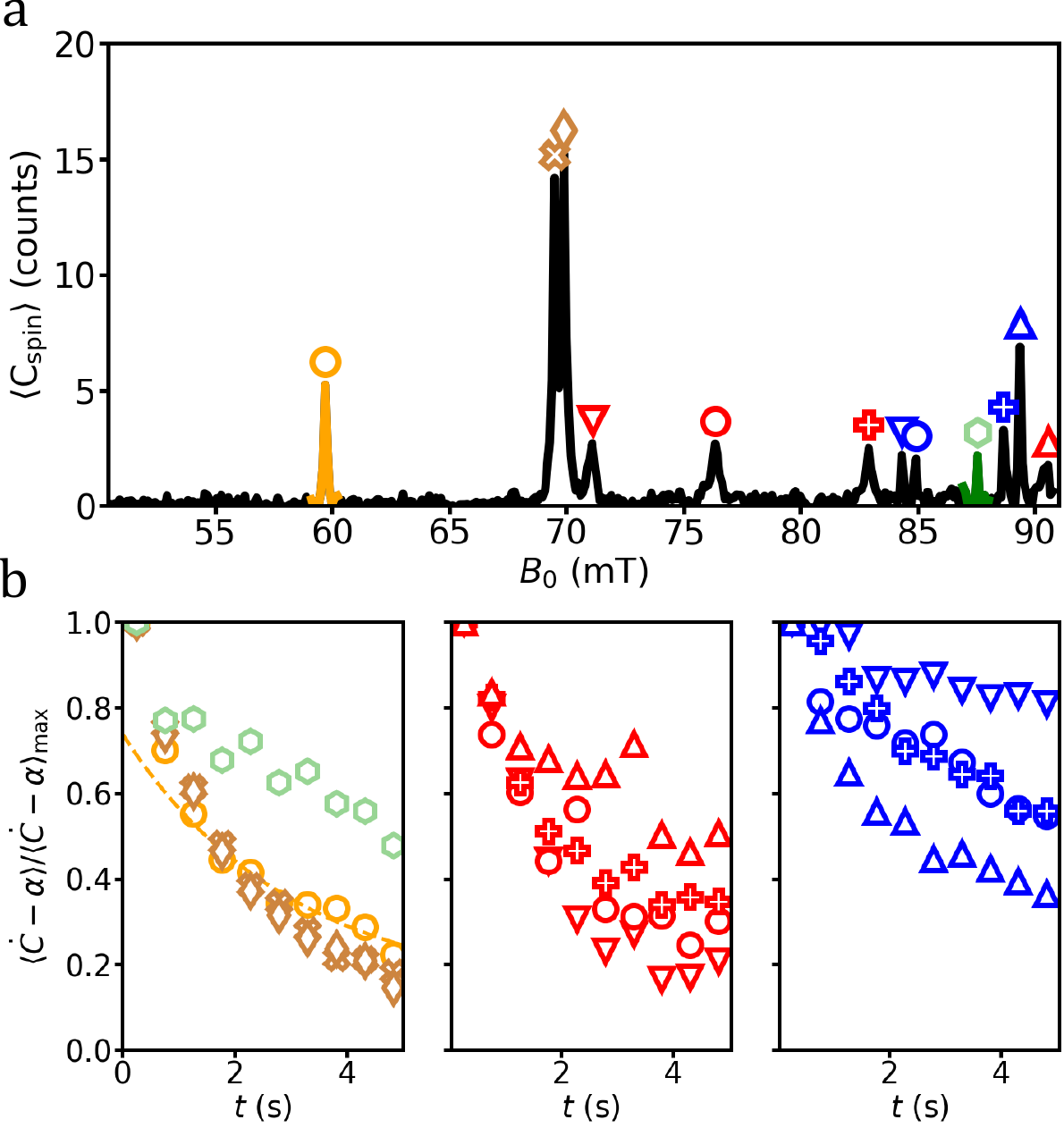}
    \caption{\label{fig6}
    \textbf{Fluorescence curves.} 
    \textbf{a}. Spectrum recorded at $\varphi = -2^\circ$ (same data as in \ref{fig5}). The labels are used to identify the resonance curves below.
    \textbf{b}, Normalized fluorescence decay curves with dark count $\alpha$ subtracted, where $\alpha = 430$ counts/s is computed using the average rate in the field range $B_0 \in [50.5: 58.5]$ mT where no peaks are visible. Left panel: fluorescence curves of $\text{Er}^{3+}$ $I=0$ transition (yellow), $^{173}\text{Yb}$ $m_I=-5/2$ transition (green), $b_1$ and $b_2$ transitions (brown). Middle panel: fluorescence decay curves of the peaks $l_i$ with $i \in [1:4]$. right panel: fluorescence decay curves of the four $\mathrm{Fe}^{3+}:\mathrm{CaWO}_4$ Kramers ground-state doublet transitions, corresponding to different orientations.
    }
\end{figure}

In addition to the mere number of counts, FD EPR spectroscopy also gives a direct access to the longitudinal relaxation of the paramagnetic impurities. In Fig.~\ref{fig6}, the fluorescence curve of each peak visible in the $\varphi = -2^{\circ}$ spectrum is shown. In these data, the measured dark count $\alpha$ was subtracted from $\langle \dot{C} \rangle$, and the resulting fluorescence signal was normalized to $1$ for convenient comparison.  All the curves show qualitatively similar behavior, decaying rapidly at short-times, followed by slower, approximately exponential decay, as expected from the discussion in Section II. This is particularly evident for the $\mathrm{Er}^{3+}$ transition, and also for the $b1$ and $b2$ transitions. The $\mathrm{Yb}^{3+}$ peak relaxes slower than $\mathrm{Er}^{3+}$. This is expected for the radiative contribution, since the gyromagnetic ratio of $\mathrm{Yb}^{3+}$ is approximately half that of $\mathrm{Er}^{3+}$, and therefore $\Gamma_R$ approximately four times lower. The data are however insufficient to determine if the observed decay is radiative or non-radiative, because the complete fluorescence curve was not measured as it exceeded the measurement time $T_{int}$. 

The four $\mathrm{Fe}^{3+}$ transitions overall show much slower relaxation, which implies that the $\mathrm{Fe}^{3+}$ spin-lattice relaxation rate $\Gamma_{NR}$ is correspondingly also much lower than the one of $\mathrm{Er}^{3+}$. This is expected for a transition metal ion that lacks the ground-state orbital moment ($L=0$, $S = 5/2$) necessary for a spin-orbit interaction that can mediate significant spin-lattice coupling and therefore relaxation. Instead, spin-lattice relaxation must be weakly mediated here by either a direct magnetic-phonon interaction, or more likely some admixture of non-zero orbital moments from higher energy states allowed by a strong crystal-field interaction, typical of transition-metal-ions in solid-state hosts.
One of those peaks shows the short-time non-exponential Purcell contribution, which is less clear in the three others, probably due to the orientation dependence of the coupling constant and therefore of the radiative relaxation rate. Our data makes it difficult to be more quantitative because the repetition time $T_{rep}$ was chosen to optimally measure the $\mathrm{Er}^{3+}$ transition and was therefore too short to properly measure $\mathrm{Fe}^{3+}$; they show nevertheless that the non-radiative (spin-lattice) relaxation rate of the ground-state Kramers doublet of $\mathrm{Fe}^{3+}$ is lower than $\sim 10^{-2} \mathrm{s}^{-1}$ (at $B_0 \sim 90$\,mT). The fluorescence curves of the four unknown $S_4$ transitions also show a fast decay (Purcell contribution), and a much slower long-time decay. The radiative rate appears to be lower for the high-field transitions, possibly indicating a lower perpendicular gyromagnetic ratio.

Overall, even though the present discussion remains qualitative, we believe that the analysis of fluorescence curves can yield useful quantitative information, which constitutes another interesting aspect of the FD-EPR method, compared to ID-EPR. 

\section{Fluorescence curves analysis}

We now provide a detailed study and quantitative analysis of the fluorescence curves, focusing on the $I=0$ $\mathrm{Er}^{3+}$ transition as a model system. By applying a static magnetic field $B_0 = 60.3$ mT at an angle $\varphi = 47^\circ$ with respect to the crystal a-axis, we resonantly excite the $\mathrm{Er}^{3+}$ I=0 transition. We then measure, using setup 1, the fluorescence curves $\langle \dot{C} \rangle (t)$ following an excitation pulse of duration $Dt$ and amplitude $\beta$ applied at $t=0$. Background subtraction is achieved by subtracting for each curve the final count rate $\langle \dot{C}(T_r) \rangle $ (averaged over the curve last $T_{\text{rep}}/6$ seconds), as described in Section III. The curves were obtained in $2$ batches : one in which the duration $Dt$ was fixed, and $\beta$ increased, and one in which $\beta$ was set to the maximal value allowed by the experimental setup and $Dt$ increased.

\begin{figure}[t!]
    \includegraphics[width=\columnwidth]{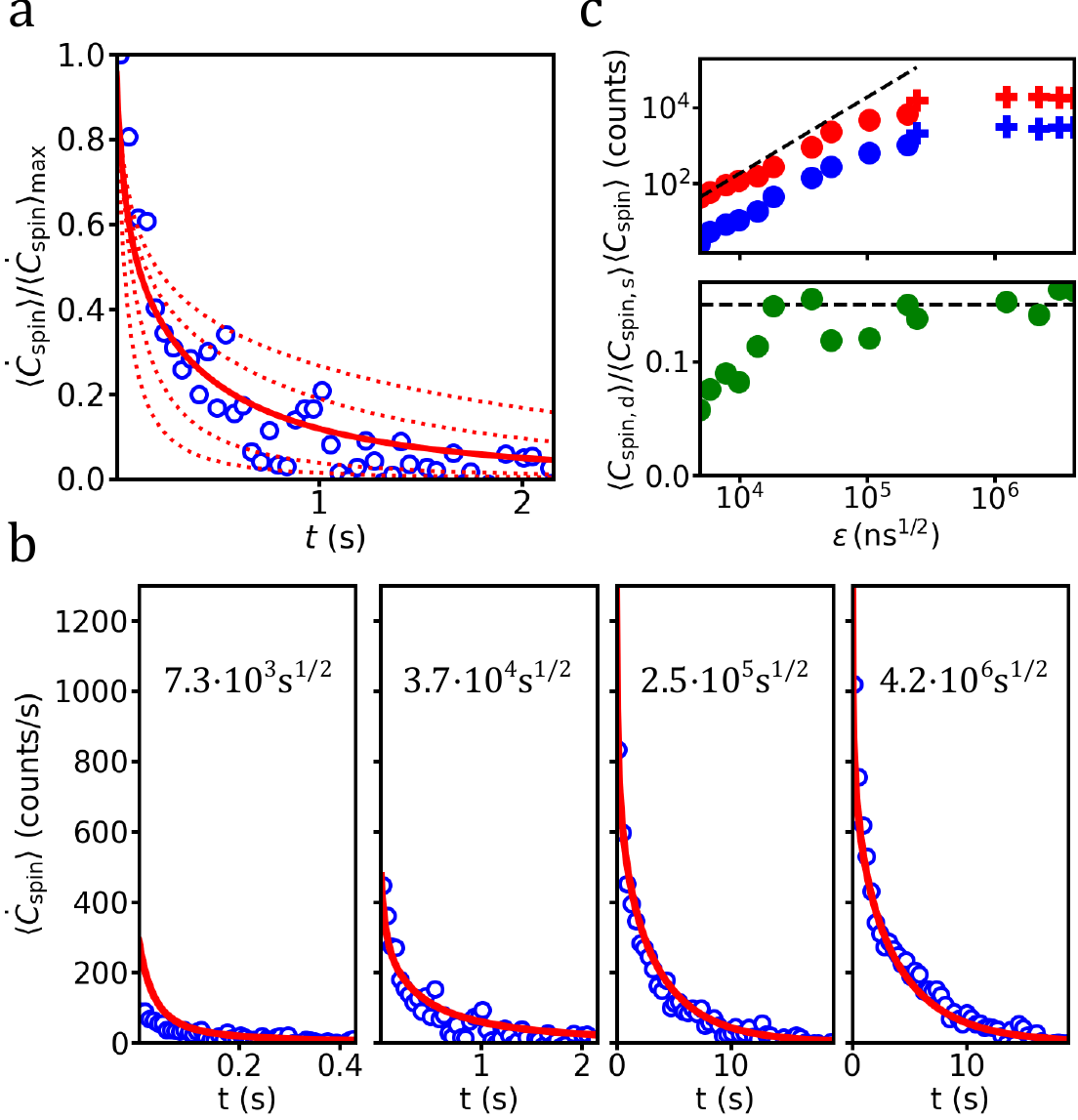}
    \caption{\label{fig7}
    \textbf{Quantitative modeling of the fluorescence decay curves.}
    \textbf{a}, normalized fluorescence curve with background subtracted 
    (open blue circles) and several simulation curves (dotted red lines) with varying excitation pulse amplitude. Best agreement is obtained with $\beta = 9.3 \, \text{ns}^{-1/2}$ (solid red line), yielding the input-line attenuation $A = 85$\,dB.
    \textbf{b}, top panel: integrated counts with background subtracted for data (blue) and simulation (red) as a function of pulse strength $\epsilon$. Dashed black line represents $\epsilon^2$ scaling. Bottom panel: green circles represent the ratio of experimentally detected $\langle C_{\mathrm{spin,d}} \rangle$ to simulated counts $\langle C_{\mathrm{spin,s}} \rangle$, as a function of excitation strength $\epsilon$ (green dots). The ratio in the high-$\epsilon$ limit is approximately constant, yielding the overall spin-to-click efficiency $\eta = 0.15$ (black dashed line).
    \textbf{c}, spin relaxation with background subtraction (open blue circles). Overlaid are simulated relaxation curves (red solid line) computed using parameters $A = 85$\,dB and $\eta = 0.15$, for four values of $\epsilon$ (values listed in each panel).
    }
\end{figure}

Prior to analyzing the data, we estimate the line attenuation $A$. For this, we consider a fluorescence curve (shown in Fig.~\ref{fig7}a) and compare it with the simulation for various values of $\beta$. The simulations (see Section II) use the experimentally-determined parameters (pulse duration $Dt$, repetition time $T_{rep}$, erbium concentration, erbium linewidth), as well as the coupling constant distribution $\rho(g_0)$ computed from the resonator geometry (see Section III), and the measured non-radiative relaxation rate $\Gamma_{NR} = 0.15 \mathrm{s}^{-1}$. The computed curves are background-subtracted and normalized in the same way as the data. The curve that fits best the data corresponds to $\beta = 9.3 \, \text{ns}^{-1/2}$, yielding a total attenuation $A = 85$\,dB, close to the value expected from the cryostat wiring. 

We then compute the simulated number of counts $\langle C_\mathrm{spin,s} \rangle$ for a given integration time $T_{int}$, and compare it to the experimental number of counts $\langle C_\mathrm{spin,d} \rangle$ over the same integration time. Both quantities are plotted in Fig.~\ref{fig7}b as a function of $\epsilon$, on a range spanning more than two orders of magnitude. The two curves demonstrate the same qualitative behavior: the number of counts increases approximately in proportion to $\sim \epsilon^2$ for small $\epsilon$, before saturating at larger values of $\epsilon$. As shown in Appendix B, the quadratic increase is due to the contribution of spins within the volume $V_R$, whereas at higher powers the contribution from spins within $V_{NR}$ is more significant, and grows only logarithmically, explaining the observed saturation. 

The experimental number of counts $\langle C_\mathrm{spin,d} \rangle$ is smaller than the simulated one $\langle C_\mathrm{spin,s} \rangle$ by an approximately constant factor, as seen in Fig.~\ref{fig7}b. More precisely, the ratio $\langle C_\mathrm{spin,d} \rangle / \langle C_\mathrm{spin,s} \rangle$ asymptotes to a value of $0.15$ in the high-$\epsilon$ limit, while decreasing in the low-$\epsilon$ limit, for reasons discussed in Section VII. Nevertheless, this ratio varies by less than a factor $2$ over nearly 3 orders of magnitude of drive amplitude $\epsilon$, supporting our interpretation that the overall efficiency is $\eta = 0.15 \pm 0.03$ for the present measurements. This value is readily explained by a combination of resonator losses, microwave transmission losses between the sample and the SMPD, and SMPD finite efficiency.

Having determined the input-line attenuation and overall efficiency $\eta$, we can quantitatively compare the measured and simulated fluorescence decay curves, over the entire experimentally achievable range of $\epsilon$. A selection of $4$ decay curves are shown in Fig.~\ref{fig7}, with their corresponding simulations. The agreement with the simulations is quantitative for values of $\epsilon$ above $\sim 3 \cdot 10^4 \mathrm{s}^{1/2}$. The decay time constant and the amplitude of the fluorescence curves are well captured by the simulations, despite their strong dependence on $\epsilon$. This validates the modeling of the fluorescence signal as the incoherent sum of the spontaneous emission of all spins, as described by Eq.~\ref{eq:fluo}. 

At low values of $\epsilon$ however, a clear discrepancy is observed. Contrary to the simulated decay curve, here the data shows a lack of contribution from the most rapidly relaxing spins (and thus, from the spins with largest coupling $g_0$). In Section VII, we present spectroscopic evidence that these spins are rather frequency-shifted by strain originating from the thermal contraction of the resonator metallic wire. 


\section{Comparison of Signal-to-noise ratios of ID-EPR and FD-EPR}

\begin{figure}[t!]
    \includegraphics[width=\columnwidth]{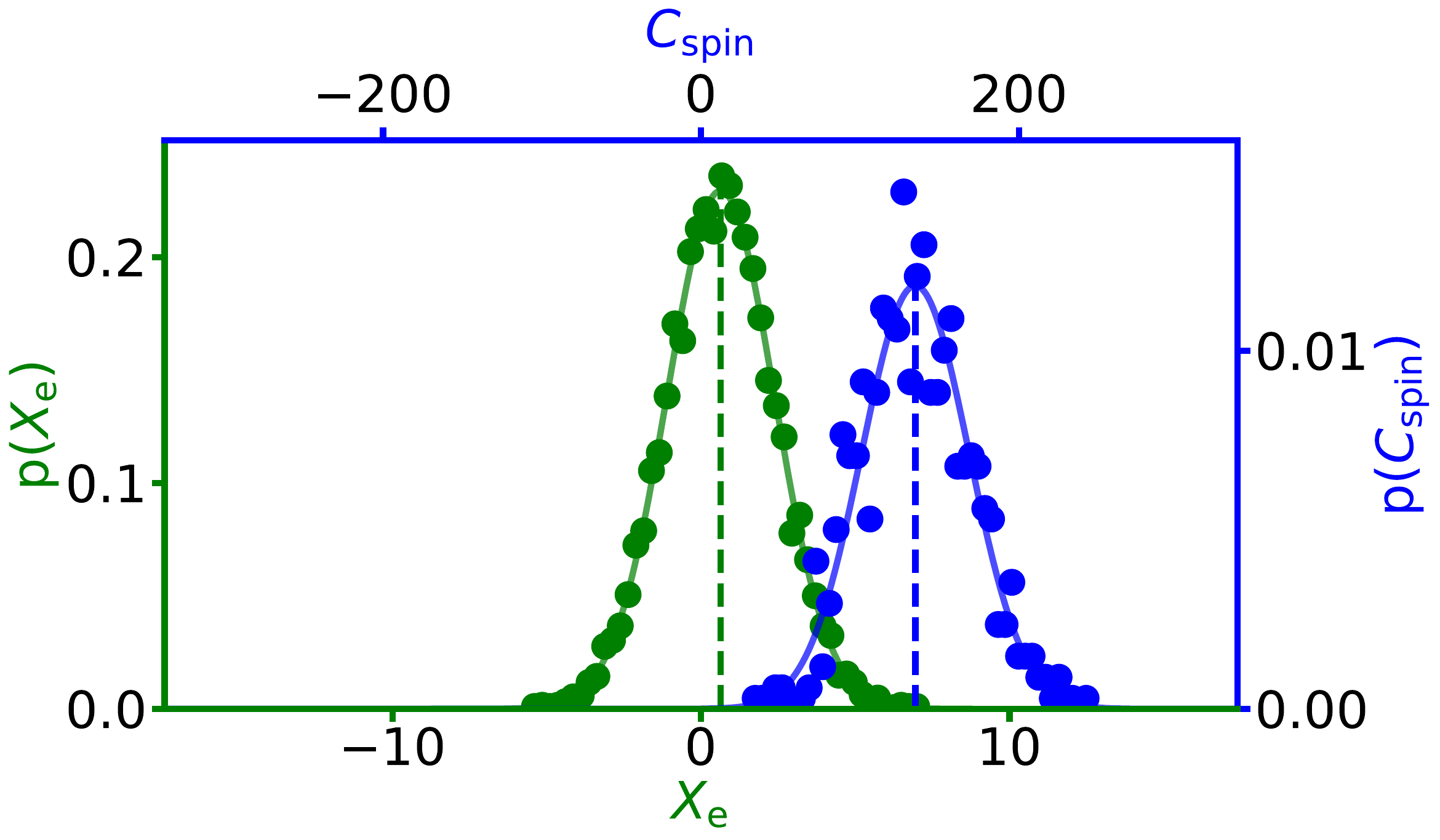}
    \caption{\label{fig8}
    \textbf{SNR comparison between ID-EPR and FD-EPR at fixed excitation power.} Green (resp. blue) circles show the measured probability of echo amplitude $X_e$ (resp. number of counts $C_{spin}$). The pulse amplitudes are $\epsilon = 3.9 \times 10^4 \, \text{ns}^{1/2}$ (resp. $\epsilon = 4.2 \times 10^4 \, \text{ns}^{1/2}$) for ID echoes (resp. for FD number of counts). ID-EPR has $\langle X_e \rangle = 0.5$ (green vertical dashed line) and $\delta X_e = 1.53$, yielding a SNR of $0.33$. FD-EPR has $\langle C_{\text{spin}} \rangle = 85$ counts (blue vertical dashed line) and $\delta C = 28$, yielding an almost order-of-magnitude larger SNR of $3$.
    }
\end{figure}


The presence of a Parametric Amplifier in the detection line of Setup $1$ enables quantitative comparison between the signal-to-noise ratio of fluorescence and inductive detection, under the same experimental conditions.

In order to compare the SNR, we repeatedly apply measurement sequences such as those described in Sections III to V, using an input pulse of amplitude $\beta$ and duration $Dt$, and compute the count probability histogram. The background rate is averaged over all the iterations at a given $\epsilon$ before subtraction, in order to minimize the added noise due to background fluctuations. $T_{rep}$ varies depending on $\epsilon$ in order for to decay to an almost constant background rate. The choice of integration time, $T_{int}$, has a strong influence on the signal SNR as longer $T_{int}$ integrates more spin signal but also more noise. Therefore, this parameter is chosen for each value of $\epsilon$ in order to maximize the SNR. Typical data are shown in Fig.~\ref{fig8}; the histogram is well fitted by a Gaussian curve of mean value $\langle C \rangle$ (the fluorescence signal) and standard deviation $\delta C$ (the noise). 

To compare the resulting SNR with that obtained from ID-EPR, we perform measurements of Hahn echoes, detected by amplification followed by demodulation, in the usual inductive-detection (ID) method, as described in Section III. The sequence consists of a first pulse of amplitude $\beta/2$ and duration $Dt$, followed after a delay $\tau = 50 \mu \mathrm{s}$ by a refocusing pulse of amplitude $\beta$ and duration $Dt$. The echo amplitude $X_e$ is then obtained by integrating the signal contained within the echo mode, which is determined by the ensemble averaged echo time trace $\langle X_e \rangle (t)$. It is converted into dimensionless unit by using the SMPD as an absolute intensity detector ~\cite{billaud_microwave_2023}. A histogram of the echo amplitude $X_e$ is shown in Fig.~\ref{fig8}. It is also well fitted by a Gaussian curve of mean value $\langle X_e \rangle$ (the echo signal) and standard deviation $\delta X_e$ (the noise). 

The two histograms obtained by fluorescence and echo detection are shown in Fig.~\ref{fig8} for similar values of $\epsilon$.  The SNR of the fluorescence signal is $\sim 10$ times larger than the echo signal at the lowest experimental excitation power, which confirms that fluorescence-detected EPR spectroscopy can be more sensitive than inductively-detected EPR at the quantum limit. In a simplified model that assumes a homogeneous distribution of relaxation rates $\Gamma_R$, and assuming that the partition noise is negligible during the fluorescence signal detection, the ratio $SNR_{FD}/SNR_{ID}$ is expected to be $\sim \sqrt{\frac{\eta \kappa}{2\alpha}}$, which amounts to $\sim 10$ for our parameters, in reasonable agreement with the measured values. In a related experiment with the same sample, the SNR of a Hahn echo was also found to be larger when detected by fluorescence than when detected inductively~\cite{billaud_microwave_2023}. The SNR gain ($\sim 1.6$) was smaller than in the present experiment due to the inhomogeneous distribution of Rabi frequencies which affects the echo pulses and particularly the 3-pulse FD echo~\cite{billaud_microwave_2023}.

\begin{figure}[t!]
    \includegraphics[width=\columnwidth]{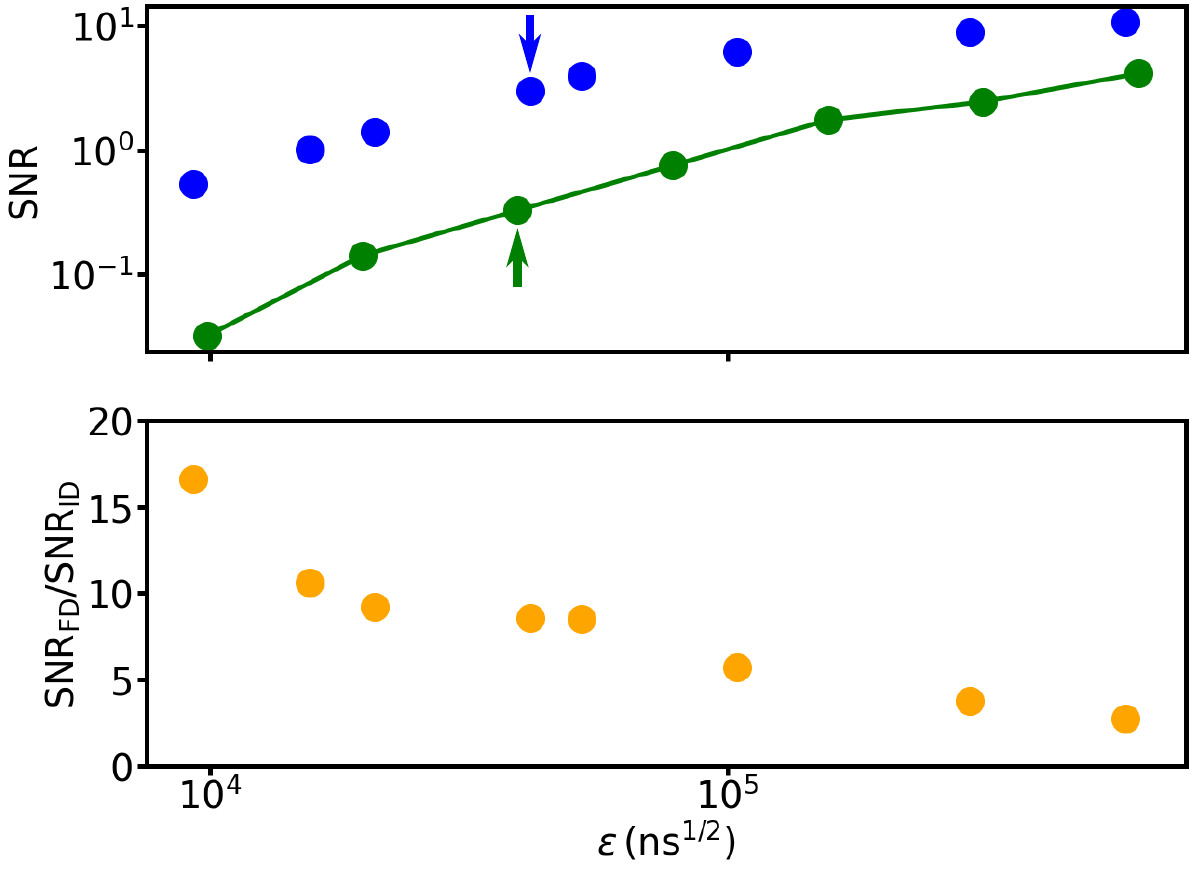}
    \caption{\label{fig9}
    \textbf{ID-EPR and FD-EPR SNR comparison as a function of excitation power.}
    Top panel: measured SNR using ID (green circles) and FD (blue circles) as a function of $\epsilon$. In order to directly compare the SNR or ID-EPR and FD-EPR at equivalent values of $\epsilon$ (see bottom panel), we linearly interpolate between ID-EPR measurements (solid green line) to infer the necessary values of ID SNR. The ID-SNR and FD-SNR measurements presented in Fig.~\ref{fig8} are highlighted with green and blue arrows, respectively. Bottom panel: yellow circles show the ratio of FD-SNR to ID-SNR as a function of $\epsilon$.
    }
\end{figure}

We then further study the SNR of fluorescence and echo detection as a function of the drive pulse amplitude $\epsilon$, and the result is presented in Fig.~\ref{fig9}. Both SNRs increase with $\epsilon$; no saturation is visible although the slope is decreasing at high $\epsilon$. The ratio of SNRs, however, decreases with increasing drive strength, which is not expected in the simplified model. Additional measurements show that this decrease results from inhomogeneity in $\Gamma_R$, causing the integration time to increase with $\epsilon$ for the fluorescence signals (as shown in Section V), whereas it is constant for ID echo detection, since the echo mode remains unchanged. Overall, these measurements show that fluorescence-detection is particularly advantageous in the limit of low-excitation power where a small number of spins are detected; in that limit, improvements in SNR of more than an order of magnitude (and thus of more than two orders of magnitude in measurement time) can be obtained.

\section{Strain shifts}

\subsection{EPR spectroscopy of strained Rare-Earth-Ions}

\begin{figure}[t!]
    \includegraphics[width=\columnwidth]{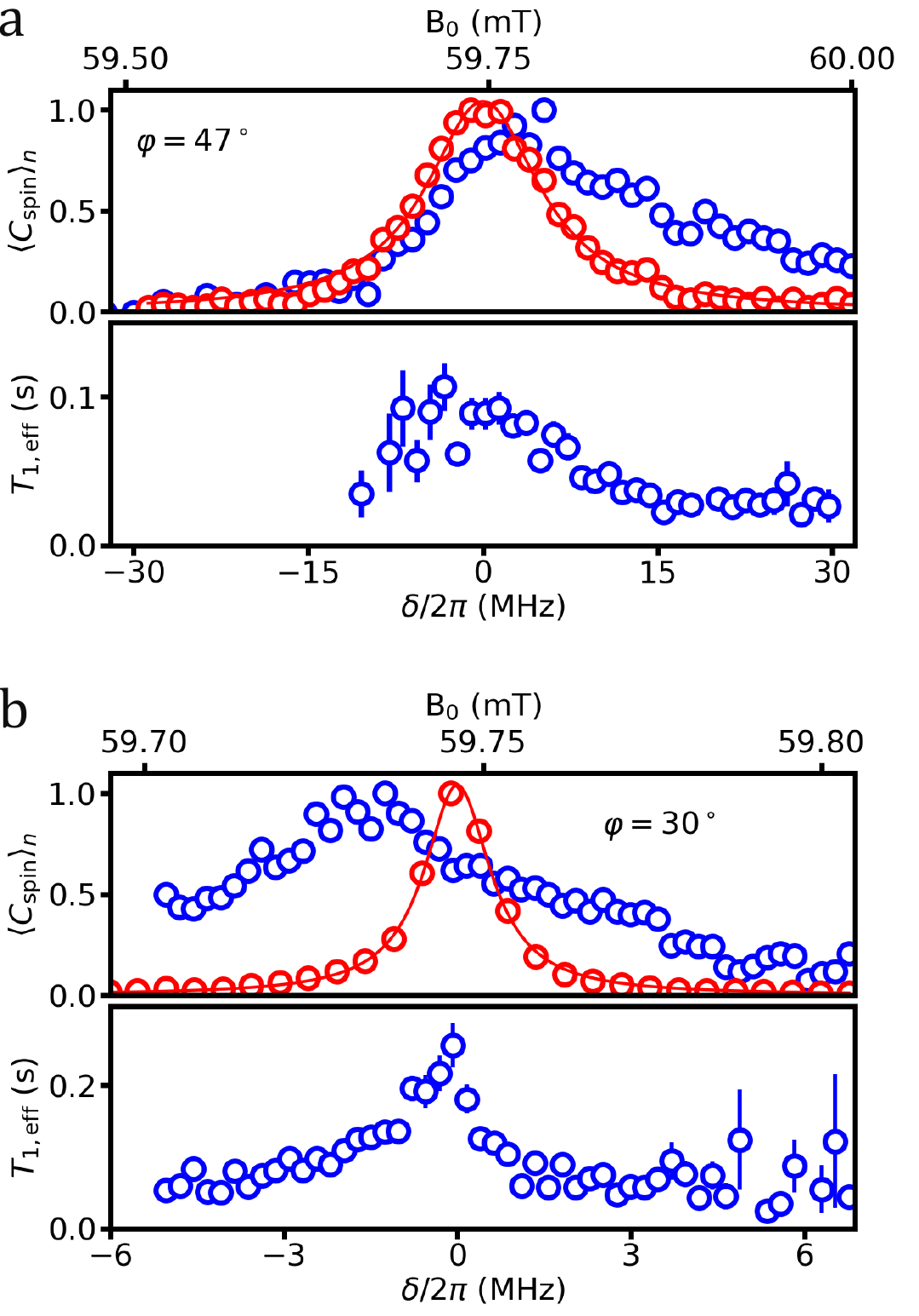}
    \caption{\label{fig10}
    \textbf{Low and high power $\mathrm{Er}^{3+}$ I=0 spectra, and relaxation times.} 
    \textbf{a} Top panel: Normalized spectrum of the $I=0$ $\mathrm{Er}^{3+}$ transition measured at $\varphi = 47 ^\circ$ using FD and setup 1. Red open circles are measured at high excitation pulse power ($\epsilon = 2.1 \times 10^5 \, \mathrm{ns}^{1/2}$, $T_{\text{int}} = 3$ s, $T_{\text{rep}} = 8.6$ s, red dots). Blue open circles are measured at low excitation strength ($\epsilon = 1.0 \times 10^4 \, \mathrm{ns}^{1/2}$, $T_{\text{int}} = 0.05$ s, $T_{\text{rep}} = 0.052$ s). The peaks of both spectra are normalised to 1 to ease comparison. The red solid line through the high-power spectrum is a Lorentzian fit yielding $\Gamma_{\text{inh}}/2\pi = 11$ MHz. 
    Bottom panel: Effective relaxation time $T_{1,eff}$ determined from the low-power spectrum, using an exponential fit to the fluorescence decay curves. The error-bars represent the standard-error in the fit. Only the values where the SNR is sufficiently large to perform the fit are plotted. 
    \textbf{b} Top panel: Normalized spectrum of the $I=0$ $\mathrm{Er}^{3+}$ transition measured at $\varphi = 30 ^\circ$ using FD and setup 2. Red open circles are measured at high excitation pulse strength ($\epsilon = 3.5 \times 10^5 \, \mathrm{ns}^{1/2}$, $T_{\text{int}} = 0.75$ s, $T_{\text{rep}} = 3$ s). Blue open circles are measured at low excitation strength ($\epsilon = 5.8 \times 10^3 \, \mathrm{ns}^{1/2}$, $T_{\text{int}} = 0.05$ s, $T_{\text{rep}} = 1$ s). The peaks of both spectra are normalised to 1 to ease comparison. The red solid line through the high-power spectrum is a Lorentzian fit yielding $\Gamma_{\text{inh}}/2\pi = 1.4$ MHz. 
    \textbf{b} Bottom panel: Effective relaxation time $T_{1,eff}$ determined from the lower-power spectrum data presented in the top panel, using an exponential fit to the fluorescence decay curves. The error-bars represent the standard-error in the fit. 
    }
\end{figure}

Thanks to the high sensitivity of fluorescence detection, it is now possible to measure EPR spectra at very low excitation powers, where only the spins closest to the wire are excited. In these conditions, we observe strong line distortions that we interpret as being caused by strain frequency shifts, which we investigate further below.

This line distorsion is evidenced in Fig.\ref{fig10}, where we compare two fluorescence-detected spectra of the erbium $I=0$ transition, both measured with the static field at an angle $\varphi = 47^\circ$, but with vastly different excitation powers. For easy comparison, the two spectra were re-normalized to $1$. The high power spectrum shows a symmetric, approximately Lorentzian lineshape, with a width $\Gamma /2\pi = 11$\,MHz, dominated by electric-field broadening as discussed in Section II. The low-power spectrum, on the other hand, shows a broader and asymmetric lineshape, with a pronounced tail on the high-field side.

\begin{figure}[t!]
    \includegraphics[width=\columnwidth]{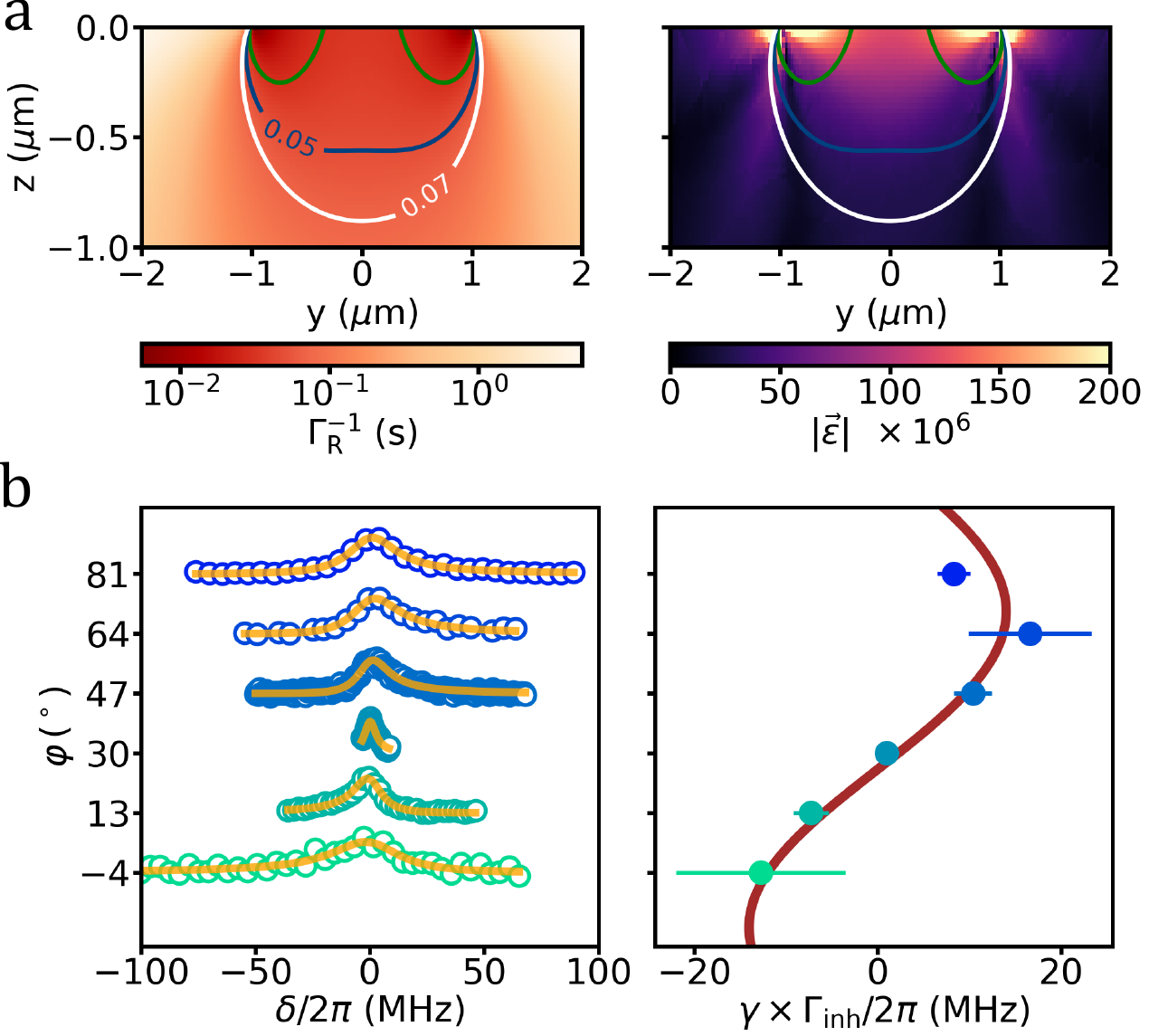}
    \caption{\label{fig11}
    \textbf{Strain shift spectroscopy.}
    \textbf{a}, Left: Surface plot of the radiative relaxation time $\Gamma_{\text{R}}^{-1}$.
    Right: Surface plot of the simulated hydrostatic strain $|\Vec{\epsilon}| = \sqrt{\epsilon_{x,x}^2 + \epsilon_{y,y}^2 + \epsilon_{z,z}^2}$. On both plots the contours show the position where $\Gamma_{\text{R}}^{-1} = 0.03$ s (green), $\Gamma_{\text{R}}^{-1} = 0.05$ s (blue), $\Gamma_{\text{R}}^{-1} = 0.07$ s (white).
    \textbf{b}, left panel: normalized spectra taken at angular increments of $17^\circ$ in $\varphi$, starting from $\varphi = -4^\circ$. The excitation pulse parameters ($\epsilon$, $T_{\text{rep}}$) vary between spectra, and were chosen to maximise signal for each specific measurement. The integration time was fixed at $T_{\text{int}} = 0.05$ s. The orange fits were performed using a skewed Lorentzian model.
    Right panel: fitted skewness parameter $\gamma \times \Gamma_{\mathrm{inh}}$ as a function of $\varphi$. The error-bars represent the standard-error in the fit. The brown line represents a fit of the transition skewness as a function of magnetic field angle $\varphi$ by a sine curve, yielding a zero-skewness angle of $26 \, ^\circ$.
    }
\end{figure}

More insight is obtained by analyzing the fluorescence curves, and fitting their effective decay time constant $T_{1,eff}$ by a single exponential. As seen in Fig.\ref{fig10}, $T_{1,eff}$ is largest on the high-power peak, and reduces by up to a factor $3$ at higher field, reaching $\sim 30$\,ms. This indicates that the high-field tail originates from spins that have radiative rates that exceed those in the bulk, and are therefore closer to the resonator wire. To be more quantitative, we simulate the theoretical $\Gamma_R$ as a function of the $\mathrm{Er}^{3+}$ spin position relative to the wire in Fig.\ref{fig11}. We see that the spins with $\sim 30$\,ms radiative lifetime are found directly underneath the wire, at depths not exceeding $500$\,nm. In this region, it is well established that significant strain is expected, due to the differential thermal contraction of the metallic wire with respect to the insulating substrate. Following the method described in reference ~\cite{pla_strain-induced_2018}, the expected strain tensor map can be computed. The hydrostatic strain profile is, for instance, shown in Fig.\ref{fig11}, and we see that the $30-50$\,ms-radiative-lifetime region indeed coincides with a region of high strain. 

Having established that the lineshape distortion is likely caused by strain, we can study its behavior as a function of $\varphi$, as outlined in Section II. In increments of $17$ degrees, starting from $\varphi = -4^\circ$, we record a spectrum of the $\mathrm{Er}^{3+}$ I=0 transition at the lowest possible power, where strain shifts are expected to clearly affect the lineshape. A waterfall plot of these spectra are shown in Fig 11 b, where the narrowest line is observed when $\varphi \sim \varphi_0$, as expected from the phenomenological reasoning put forward in Section II. Moreover, the asymmetry of the line appears to be mirrored when $\varphi$ is tuned across the $\varphi_0$ value. This confirms, at least qualitatively, the reasoning presented in Section II, and in particular that spins located at inversion-related sites subject to the same strain tensor exhibit the same frequency shift.

\subsection{Coherent oscillations}

As was shown in the previous section, strain shifts introduce a correlation between the Larmor frequency of an electronic spin and its position in space relative to the wire. Therefore, it raises the possibility of addressing spins selectively within a spatial region simply by tuning $B_0$. This phenomenon was used recently to measure a spatial map of the coherence time of bismuth donors in silicon~\cite{ranjan_spatially-resolved_2021}. Here, we confirm this spatial selectivity through the observation of coherent oscillations in the ensemble spin-resonance in the highly strain-shifted region, which is an indication that the measured spins are subjected to the same oscillating magnetic field amplitude and therefore in a region of space small compared to the amplitude gradient of $B_1$. By contrast, as explained in Section II, Rabi oscillations cannot be observed for the bulk of the unstrained spins, whose transition frequencies lie close to the centre of the absorption peak, because of the large spread of Rabi angles throughout the ensemble.

\begin{figure}[t!]
    \includegraphics[width=\columnwidth]{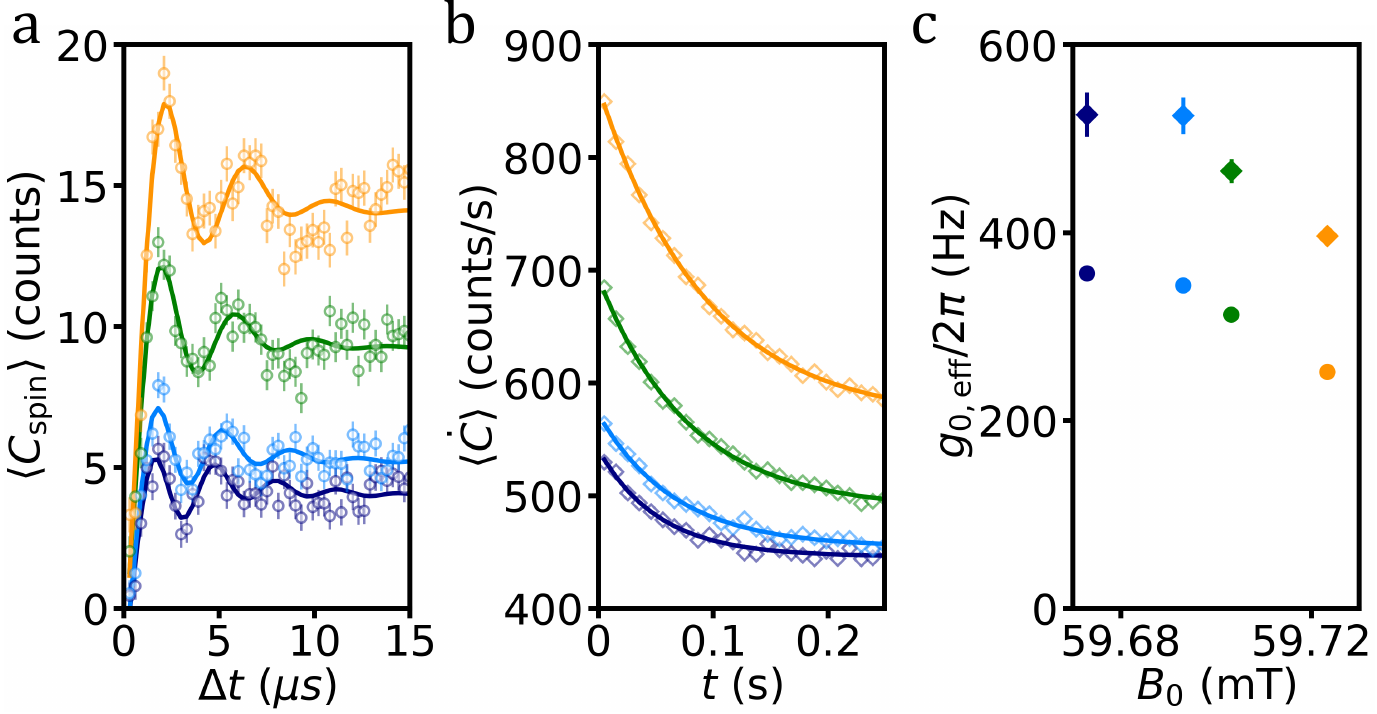}
    \caption{\label{fig12}
    \textbf{Rabi oscillations, decay curves and inferred spin-resonator coupling.}
    Data measured at four field amplitudes for a magnetic field angle $\varphi = 30^\circ$, from $B_0 = 59.673$ mT (dark blue) to $B_0 = 59.723$ (orange) in steps of 0.01 mT, using parameters $\beta = 11.5 \, \text{ns}^{-1/2}$, $T_{\text{int}} = 0.075$ s and $T_{\text{rep}} = 0.25$ s.
    \textbf{a}, Measured average number of counts $\langle C_{spin} \rangle$ (open circles) as a function of pulse duration $\Delta t$. Solid lines are a phenomenological fit to the data (see main text).
    \textbf{b}, Fluorescence decay curves taken with the same experimental paramaters averaged over all $\Delta t$ (open diamonds) and their corresponding exponential fits (solid lines).
    \textbf{c}, Effective spin-resonator coupling $g_{\text{0,eff}}$ determined from the Rabi oscillation data (open circles) and from the fluorescence curves (hexagons). The error bars are $1\sigma$ deduced from the fits.
    }
\end{figure}

In Fig.~\ref{fig12}, the background-subtracted integrated number of counts $\langle C_{spin} \rangle$ is shown as a function of the excitation pulse length $Dt$, for various values of the magnetic field $B_0$ applied at an angle $\varphi = \varphi_0 = 30^\circ$, on the low-field tail of the erbium transition. Damped oscillations are observed in all the curves, with an increasing oscillation frequency for decreasing field values. This correlation can be understood in terms of the degree of strain affecting spins in close proximity to the inductive resonator wire; the strain yields larger frequency shifts (i.e: resonant-field shifts) for increasingly proximal spins, while simultaneously increasing the spin-resonator coupling strength (i.e: Rabi frequency) for spins within the same proximity. Moreover, the oscillations are superposed on a background signal which strongly increases as the field is tuned towards the centre of the absorption peak, and which likely comes from the bulk erbium spins that are unaffected by strain. Note that the oscillation amplitude ($\sim 3 - 5$ counts on average) indicates that this signal comes from $\sim 20-35$ spins.

For a more quantitative analysis, we fit each curve by a damped sinusoid together with a decaying exponential, $f(\Delta t) = A \sin^2  \frac{ \Omega_{R} \Delta t}{2} \mathrm{e}^{- \Delta t/T_{c,1}} + B(1 - \mathrm{e}^{- \Delta t/T_{c,2}})$. From the Rabi frequency, we obtain the effective coupling constant $g_{0,eff} = \Omega_R / 2 \sqrt{\bar{n}}$, $\bar{n}$ being the mean photon number in the cavity induced by the drive pulse, which can be determined using the input line attenuation calibration described in Section V. We see in Fig.\ref{fig12}c that $g_{0,eff}$ is larger for the lower values of $B_0$, corresponding to the most strain-shifted spins. The measured coupling at the lowest values of $B_0$ is consistent with spins located directly below the wire, at an approximate subsurface depth of $500$\,nm. 

For consistency, we also measure $g_{0,eff}$ using an independent method: at each value of magnetic field, we fit the fluorescence curves obtained by averaging all the Rabi oscillations (see Fig.~\ref{fig12}b) with an exponential decay, yielding the radiative rate $\Gamma_{R}$ from which we determine the coupling constant using the Purcell formula for spins at resonance $g_{0,eff} = \sqrt{\kappa \Gamma_R} /2$. In Fig.\ref{fig12}c, we see that both methods give similar results. In particular, the decrease in coupling at higher fields is inferred both by a lower Rabi oscillation frequency and by a longer fluorescence relaxation time. The $30\%$ discrepancy between the value obtained from the Rabi and from the $T_1$ measurements was reproduced in simulations; it is due to the use of simple formulas which are valid only for spins strictly at resonance, whereas slightly detuned spins (but still within the resonator linewidth) contribute also to the signal. Theses detuned spins introduce a bias when comparing the coupling constants since they undergo faster Rabi oscillation but slower radiative relaxation than the resonant spins.

\begin{figure}[t!]
    \includegraphics[width=\columnwidth]{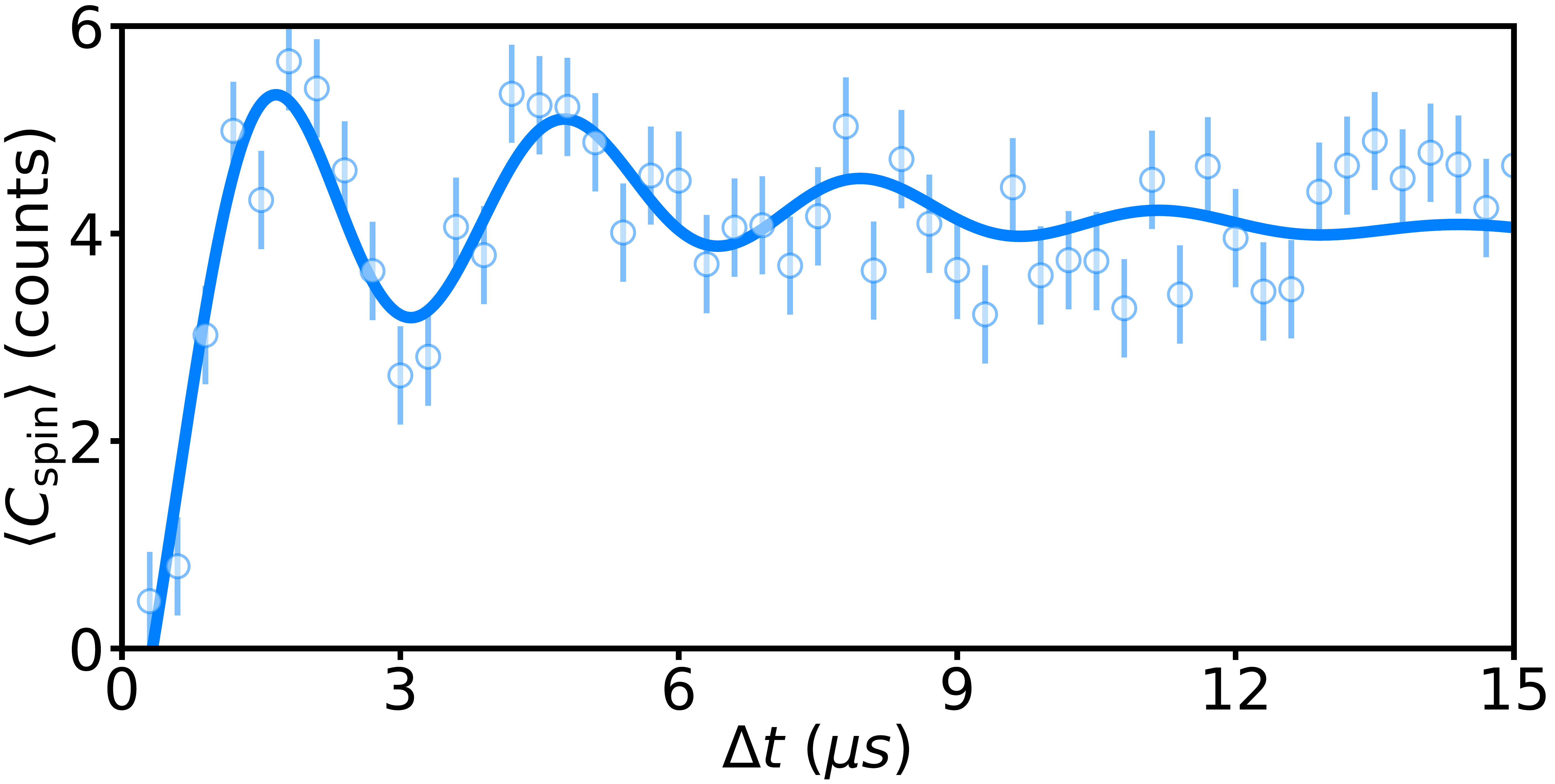}
    \caption{\label{fig13}
    \textbf{Comparison of measured and simulated Rabi oscillations}
    Measured (circles) and simulated (solid line) Rabi oscillations at $B_0 = 59.693$\,mT. Simulations include the hyperfine coupling to the $^{183}W$ nuclear-spin bath, as well as the inhomogeneous Larmor frequency distribution of the erbium electron spins. 
    }
\end{figure}

It is possible to further investigate the source of damping in the coherent oscillations observed in Fig.~\ref{fig12}a. One source is the hyperfine coupling of the erbium ions with the ligand $^{183}\mathrm{W}$ nuclear spins that surround them. To test this hypothesis, we simulate the dynamics of an erbium electron-spin coupled to $3$ $^{183}\mathrm{W}$ nuclei and driven with an excitation pulse $\alpha(t)$, using the QuTiP package. In this simulation, the hyperfine Hamiltonian is considered in the secular approximation, $H = S_z (A I_z + B I_x)$, with $I$ the nuclear spin operator, and $A$ and $B$ the isotropic and anisotropic hyperfine constants, computed for each lattice site using the point dipole approximation. Also, because only a finite number of tungsten occupation sites can be considered, the computation was made under the approximation that only the 15 nearest sites can be occupied. The nuclei occupying these sites have the strongest hyperfine constants among all and therefore lead to most of the effect. Under this assumption, and given the $14\%$ natural abundance of $^{183}\mathrm{W}$ in $\mathrm{CaWO}_4$, the probability for the erbium ions to be coupled to no more than 3 nuclear spins at the same time is 86 $\%$. After averaging the results for the $576$ total different site configurations, and also averaging the curves for spins whose frequency lies within the cavity linewidth $\kappa$, we obtain the solid red curve shown in Fig.~\ref{fig13}. It is in quantitative agreement with the data in the most highly strained region. This indicates that the dominant mechanism of the oscillation decay is the magnetic dipole-coupling of the erbium electron-spin with its nuclear spin environment (which is also causing ESEEM in a Hahn echo measurement), and that residual $g_0$ inhomogeneity amongst the strained ensemble is in fact negligible. 

Our observation of strain shifts in a REI-doped crystal provides motivation for theoretical modeling of these effects. Indeed, strain induced by thin-film circuits can be simulated with high precision, therefore, it should be possible to quantitatively model the spectra presented in Fig.~\ref{fig11}, as was done for donors in silicon~\cite{pla_strain-induced_2018,ranjan_spatially-resolved_2021}. This would, however, require a quantitative model of the dependence of the gyromagnetic tensor on the strain field, in line with the results achieved in a simpler system comprised of donors in silicon~\cite{mansir_linear_2018}.

\section{Conclusion}

We have applied the EPR microwave fluorescence detection method to a crystal of scheelite, homogeneously doped with a variety of paramagnetic ions including rare-earth-ions. We provided a quantitative analysis of the fluorescence decay curves and of their dependence on the drive pulse power in the case of erbium ions, thus validating our model to treat the emission of each spin independently. We compared the SNR of fluorescence and inductive-detection EPR spectroscopy, and found that fluorescence detection has over an order of magnitude greater sensitivity than inductive-detection in the low-power regime, which confirms the potential of fluorescence detection as a viable method for detecting small spin ensembles. Finally, we used the high sensitivity of fluorescence detection EPR spectroscopy to measure the spectral shift of spins that are located in a region of the substrate strained by the thermal contractions of the metallic thin-film from which the resonator is patterned. 

In conjunction with recent work on fluorescence-detected echo signals~\cite{billaud_microwave_2023}, these results show that FD-EPR can be considered a truly practical spectroscopy method. This detection method has also been shown recently to reach single-spin sensitivity~\cite{wang_single-electron_2023}. An additional advantage of this detection method is the ability study the ensemble-decay of the fluorescence signal in real time. There exists no equivalent time-series data that can be recorded from a single-shot measurement of ID-EPR, and here we have utilised it to study and elucidate multiple spin-relaxation channels of the paramagnetic impurities present in our sample; directly observing spin-lattice relaxation and inferring the relaxation rate. By giving access to unprecedented spin detection sensitivity, the method allows for the study of very small ensembles, at the sub-micron scale. Collectively, these measurements call for further theoretical investigation of strain-coupling in REI doped materials, in order to improve our understanding of the measured spectra of nanoscale spin ensembles.  

\section{Appendix A}

In order to identify the origin of certain signals, a study using conventional EPR (Electron Paramagnetic Resonance) was conducted on a bulk crystal from the same batch. This crystal, measuring approximately 3x3x2 mm$^3$ (mass = 112 mg), was placed in the center of a standard X-band cavity (9.62 GHz) used in a Bruker EMX spectrometer. The sample was mounted on a suprasil sample holder and secured with Apiezon N grease. Measurements were performed between 7 K and 13 K, adjusted depending on the ion being identified, as some ions exhibit varying relaxation properties. Angle measurements were taken with the c-axis aligned either parallel or perpendicular to the rotation axis. We used an automatic goniometer with an angular uncertainty of about 1$^\circ$. Not all of the data are presented here, but they are available in a data repository.

Like other ions, \ce{Fe^{3+}} is present in small quantities (a few tens of ppb) but it is highly visible due to the sharpness of its characteristic line at g=4.3. \ce{CaWO4}:\ce{Fe^{3+}} has been extensively studied by EPR in the past, and we have access to highly accurate data. In its S4 position, \ce{Fe^{3+}} does not exhibit EPR signal. But when charge compensation occurs, the local symmetry changes and the system becomes orthorhombic. Claridge et al.  reported three Kramers' doublets (see table \ref{tab:Fe}).

\begin{table*}[tbh!]
\centering
\begin{tabular}{|c|c|c|c|c|}
    \hline {$\begin{array}{l}\text { Kramers' } \\
    \text { doublet }\end{array}$} & {$k$} & {$\begin{array}{c}\text { Principal } \\
    g \text {-value, } \\
    g_k\end{array}$} & \multicolumn{2}{|c|}{ Principal directions} \\
    \hline Nomenclature & & & $\theta_k(\mathrm{deg})$ & $\phi_k(\mathrm{deg})$ \\
    \hline Upper & $\begin{array}{l}1 \\
    2 \\
    3\end{array}$ & $\begin{array}{l}9.671(18) \\
    0.986(12) \\
    0.592(80)\end{array}$ & $\begin{array}{c}123.2(2) \\
    69(2) \\
    140(2)\end{array}$ & $\begin{array}{l}40.5(2) \\
    324(2) \\
    260(4)\end{array}$ \\
    \hline Middle & $\begin{array}{l}1 \\
    2 \\
    3\end{array}$ & $\begin{array}{l}4.300(1) \\
    4.289(1) \\
    4.288(1)\end{array}$ & $\begin{array}{c}40(3) \\
    130(7) \\
    92(94)\end{array}$ & $\begin{array}{c}3(6) \\
    7(188) \\
    275(107)\end{array}$ \\
    \hline Lower & $\begin{array}{l}1 \\
    2 \\
    3\end{array}$ & $\begin{array}{l}9.670(9) \\
    1.009(5) \\
    0.488(44)\end{array}$ & $\begin{array}{l}42.4(1) \\
    113(1) \\
    123(1)\end{array}$ & $\begin{array}{c}354.8(2) \\
    57(1) \\
    310(1)\end{array}$ \\
    \hline
\end{tabular}
\caption{\textbf{Principal g-values and directions of the three Kramers doublets of \ce{CaWO4}:\ce{Fe^{3+}}.} The values are taken from  from \cite{claridge_x-band_1997}.}
 \label{tab:Fe}
\end{table*}

\begin{figure}
    \centering
    \includegraphics[width=1\linewidth]{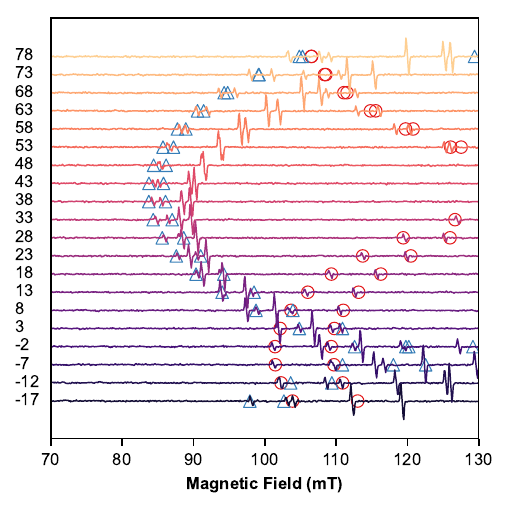}
    \caption{Angular dependence of \ce{CaWO4} EPR spectra measured at 13K. Red circles and blue triangles are the resonance fields expected using the parameters "Lower" and "Upper" doublets respectively (see table \ref{tab:Fe} \cite{claridge_x-band_1997}).}
    \label{fig:Fe}
\end{figure}

While "Middle" doublet is well identified, "Upper" and "Lower" were arbitrary chosen. At the temperature of the study (T=10K), all levels are populated and it is impossible to distinguish "Upper" and "Lower" doublet.  Fig. \ref{fig:Fe} shows the angular dependence of EPR spectra measured in the ab plane at 13K. The chosen field range approximately corresponds to that used in the measurements of the main document. As in the main document, the angle corresponds to the angle between the magnetic field and the crystal's $a$ (or $b$) axis. In order to identify the origin of the peaks, we used the parameters from Table \ref{tab:Fe}. The only slightly adjustable parameter is the crystal's disorientation, estimated to be about 2$^\circ$.

The blue triangles (red circles) correspond to the "upper" ("lower") parameters. It is important to note that only the upper series is visible in Figure \ref{fig:Fe}. At 10~mK, only the ground state is populated, allowing for a clear identification of the fundamental state and the excited states. The measurement at 10~mK proves that the doublet referred to as "upper" is actually the ground state. Only an ultra-low temperature measurement can make this distinction.

\begin{figure}
    \centering
    \includegraphics[width=1\linewidth]{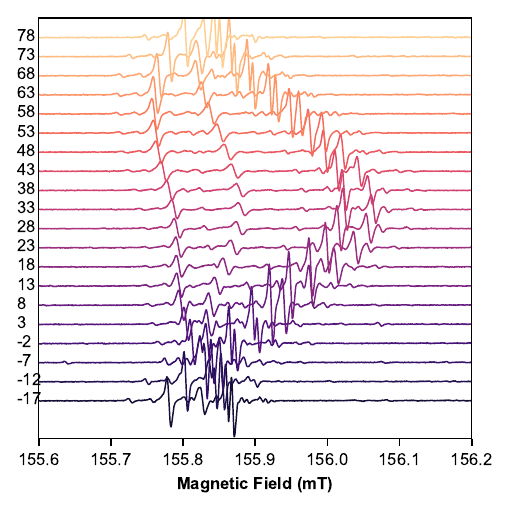}
    \caption{Angular dependence of \ce{CaWO4} near the quasi-isotropic lines (label "middle" in \ref{tab:Fe}. The resonance field change by only 0.5mT. The four main lines correspond to the 4 unquivalent position in the charge compensated sites. The small satellites are due to the superhyperfin interaction between \ce{Fe^{3+}} and \ce{^{183}W}}
    \label{fig:FeMid}
\end{figure}

The figure \ref{fig:FeMid} shows the angular dependence of the EPR lines of \ce{CaWO4} in the field range corresponding to the quasi-isotropic lines (g=4.3). Over the entire angular variation, the resonance field only varies by 0.5 mT. Four main lines can be distinguished, corresponding to the four non-equivalent charge compensation sites. The lines are very narrow, allowing the observation of satellites caused by the superhyperfine interaction between \ce{Fe^{3+}} and \ce{^{183}W}.

\begin{figure}
    \centering
    \includegraphics[width=1\linewidth]{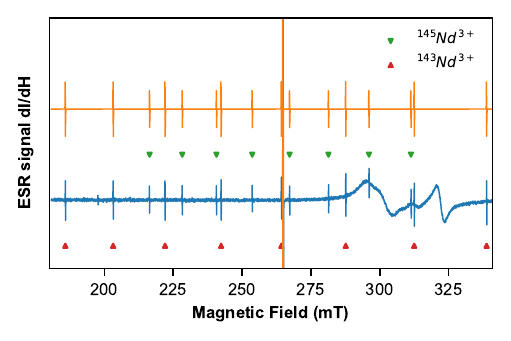}
    \caption{EPR spectrum of \ce{CaWO4} recorded at 13K and $f=9.38$~GHz with the static field perpendicular to the $c$ axis. The blue line is the experimental data while the orange line is theoretical spectrum calculated using the parameter in the text. The triangles help to identify the contribution of each isotopes.}
    \label{fig:Nd}
\end{figure}

Another series of lines observed at T=13K is also reported in Fig. \ref{fig:Nd}. The lines are very sharp and a particular attention has been done to avoid overmodulation ($h_{mod}$=0.5G). The measurement shows an intense central line surrounded by several smaller lines. The position of these lines is independent of the orientation in the ab plane and is therefore attributed to an ion in the S4 position. The position of the central line at g=2.531 is characteristic of the Nd$^{3+}$ ion in \ce{CaWO4} \cite{garrett_pmr_1964}. This ion has 7 natural isotopes, $^{142}$Nd, $^{144}$Nd, $^{146}$Nd, $^{148}$Nd, and $^{150}$Nd with $I=0$ (79.54\% abundance), which contribute to the signal of the central line, as well as 2 isotopes with nuclear spin $I=7/2$, $^{143}$Nd (12.17\%) and $^{145}$Nd (8.29\%), which are the satellite lines. The small difference in natural abundance allows them to be differentiated. To our knowledge, the hyperfin coupling has not been reported in the literature, and we obtain it here: $^{143}A_\|=606$~MHz, $^{143}A_\perp=773$~MHz, $^{145}A_\|=376$~MHz and $^{143}A_\perp=480$~MHz with an error less than 2MHz. The ratios $^{143}A_\|/^{145}A_\|\approx^{143}A_\perp/^{145}A_\perp\approx  ^{143}\mu/^{145}\mu$ confirm that the satellite line comes from Nd$^{3+}$ ions.
Using the hyperfin constants we just estimated, we can calculate the resonance fields of \ce{CaWO4}:Nd$^{3+}$ at 7~GHz, the frequency used in the main text. In the range of field accessible in the low temperature setup, only one resonance field is visible at $H_r=117.5$~mT corresponding to the line $m_I=7/2$ of $^{143}$Nd$^{3+}$.

\section{Appendix B}

Section V presents fluorescence curves measurements on the $\mathrm{Er}^{3+}$ $I=0$ transition as a function of the excitation pulse strength $\epsilon$. The number of counts was observed to increase approximately as $\sim \epsilon^2$ at low-$\epsilon$ and saturate at large values, as shown in Fig.~\ref{fig7}b. This Appendix attempts to reproduce analytically this scaling. 

The number of counts is $\langle C_{spin} \rangle = \int_{0}^{T_{int}} |\alpha_{out}|^2 (t) dt$. Using Eq.~\ref{eq:fluo}, and neglecting inhomogeneous broadening, we get that 

\begin{equation}
     |\alpha_{out}|^2 (t) = \eta \int_0^\infty dg_0  \rho(g_0) p_\epsilon(g_0) \Gamma_R(g_0) \exp^{- \Gamma_1 t},    
\end{equation}

\noindent where $p_\epsilon(g_0) = \sin^2[\psi(\epsilon)]$ is the excitation probability of spins having a coupling constant $g_0$ under a drive strength $\epsilon$, with $\psi(\epsilon) = 2 g_0 \epsilon \sqrt{\kappa_c}/\kappa $ their corresponding Rabi rotation angle. To proceed, we define a limit coupling constant $g_{0,lim}$ such that $4 g_{0,lim}^2/\kappa = \Gamma_{NR} $. We separate the contributions from spins in the Purcell volume $V_R$ (which have a coupling constant $g_0 \geq g_{0,lim}$), from those in the non-radiative volume $V_{NR}$ ($g_0 \leq g_{0,lim}$). We also assume that the integration time $T_{int}$ is inifinitely long. In the Purcell volume where $\Gamma_1 = \Gamma_R$, we get that

\begin{equation}
    \langle C_{spin,R} \rangle = \eta \int_{g_{0,lim}}^\infty dg_0  \rho(g_0) p_\epsilon(g_0).
\end{equation}

In the non-radiative volume where $\Gamma_1 = \Gamma_{NR}$, we obtain

\begin{equation}
    \langle C_{spin,NR} \rangle = \eta \int_0^{g_{0,lim}} dg_0  \rho(g_0) p_\epsilon(g_0) \frac{\Gamma_R(g_0)}{\Gamma_{NR}}.
\end{equation}

We now need to specify the coupling constant distribution. In the case of an infinitely narrow wire coupled to an infinitely large sample, $\rho(g_0) = \bar{g_0}^2/g_0^3$~\cite{ranjan_pulsed_2020}, where $\bar{g_0} = \sqrt{\langle g_0 \rangle g_{0,min}}$. Because our wire has a finite width, and our sample has a finite size, we introduce a lower coupling $g_{0,min}$ and upper coupling $g_{0,max}$, with $g_{0,min} \ll g_{0,lim} \ll g_{0,max}$. We thus obtain the following expressions

\begin{equation}\label{eq:CRg0}
    \langle C_{spin,R} \rangle = \eta \bar{g_0}^2 \int_{g_{0,lim}}^{g_{0,max}} dg_0 \frac{\sin^2[\psi(\epsilon)]}{g_0^3},
\end{equation}

and 

\begin{equation}\label{eq:CNRg0}
    \langle C_{spin,NR} \rangle = 4 \frac{\eta \bar{g_0}^2}{\kappa \Gamma_{NR}} \int_{g_{0,min}}^{g_{0,lim}} dg_0 \frac{\sin^2[\psi(\epsilon)]}{g_0}.
\end{equation}

These expressions can be rewritten using the Rabi angle as the integrated variable. A physically relevant quantity is $\psi_{lim} = 2 g_{0,lim} \epsilon \sqrt{\kappa_c}/\kappa$, the Rabi rotation angle of spins at the boundary between $V_R$ and $V_{NR}$. The expressions above become

\begin{equation}\label{eq:CRpsi}
    \langle C_{spin,R} \rangle = \frac{4 \eta \epsilon^2 \bar{g_0}^2 \kappa_C}{\kappa^2} \int_{\psi_{lim}}^{\psi_{max}} d\psi \frac{\sin^2\psi}{\psi^3},
\end{equation}

and 

\begin{equation}\label{eq:CNRpsi}
    \langle C_{spin,NR} \rangle = 4 \frac{\eta \bar{g_0}^2}{\kappa \Gamma_{NR}} \int_{\psi_{min}}^{\psi_{lim}} d\psi \frac{\sin^2\psi}{\psi}.
\end{equation}

We can now investigate how $C_{spin}$ scales with $\epsilon$ in the low- and high-$\epsilon$ limits. For the low-$\epsilon$ limit, we consider that all spins perform Rabi rotation angles much smaller than $\pi$ (i.e., $\psi_{max} \ll \pi$), such that $\sin^2 \psi \sim \psi^2$. From Eq.~\ref{eq:CRg0}, we get

\begin{equation}
    \langle C_{spin,R} \rangle \approx \frac{4 \eta  \bar{g_0}^2 \kappa_C}{\kappa^2} \mathrm{Log}\frac{g_{0,max}}{g_{0,lim}} \times \epsilon^2.
\end{equation}

For the non-radiative contribution, using the fact that $\int_{\psi_{min}}^{\psi_{lim}} d\psi \frac{\sin^2\psi}{\psi} \sim \psi_{lim}^2/2$ when $\psi_{lim} \ll \pi$, we get from Eq.~\ref{eq:CNRpsi}:

\begin{equation}
    \langle C_{spin,NR} \rangle \approx 4 \frac{\eta \bar{g_0}^2 \kappa_C}{\kappa^2} \times \epsilon^2.
\end{equation}

\noindent Since both $\langle C_{spin,R} \rangle$ and $\langle C_{spin,NR} \rangle$ scale as $\epsilon^2$, their sum $\langle C_{spin} \rangle$ also does, which accounts for the low-$\epsilon$ regime observed in Fig.~\ref{fig7}b.

For the high-$\epsilon$ limit, we consider the condition $\psi_{lim} \gg \pi$, which implies that all spins in the Purcell volume undergo a Rabi rotation angle much larger than $\pi$. In that limit, we can approximate $\sin^2 \psi$ as being equal to $1/2$ in $V_R$. From Eq.~\ref{eq:CRg0} it is clear that the radiative contribution $\langle C_{spin,R} \rangle$ is now independent from $\epsilon$. The non-radiative contribution on the other hand is proportional to $\int_0^{\psi_{lim}} d\psi \frac{\sin^2\psi}{\psi}$ (Eq.~\ref{eq:CNRpsi}), which scales like $\mathrm{Log}\, \epsilon$ in the high-$\psi_{lim}$ limit. We thus deduce that the total number of counts, $\langle C_{spin} \rangle$, increases logarithmically with the drive strength $\epsilon$ in the high-$\psi_{lim}$ limit, as observed in the measurements and simulations.

Overall, this analysis allows us to propose a physical interpretation of the change of scaling observed in Fig.~\ref{fig7}b. This change occurs when the drive strength is such that the spins located at the boundary between $V_R$ and $V_{NR}$ undergo a rotation angle of order $\pi$.

\bibliography{SpectroFluo}

\end{document}